\documentclass[final,5p,times,twocolumn,12pt,sort&compress]{elsarticle}
\usepackage{soul}
\usepackage{lineno,hyperref}
\usepackage{txfonts}
\usepackage{multirow}
\usepackage{graphicx}
\usepackage{dcolumn}
\usepackage{array}
\usepackage{color}
\usepackage{ulem}
\usepackage{CJK}
\usepackage{multicol}
\usepackage{amssymb}
\usepackage{amsmath}
\usepackage{type1cm}
\usepackage{caption}
\usepackage{natbib}
\usepackage{lineno,hyperref}
\usepackage{comment}
\usepackage{longtable}
\usepackage{bm}
\usepackage{ctable}
\usepackage{siunitx}
\modulolinenumbers[5]

\onecolumn

\journal{JQSRT, prepared by using elsarticle.cls}

\begin{document}
\begin{frontmatter}
\title{Benchmarking calculations with spectroscopic accuracy of level energies and wavelengths in \mbox{W LVII} - \mbox{W LXII} tungsten ions}

\author[fd]{Chun Yu Zhang}
\author[hb]{Kai Wang\corref{kw}}
\author[fd]{Ran Si\corref{rs}}
\author[bls]{Michel Godefroid}
\author[rd]{Per~J{\"o}nsson}
\author[fd]{Jun Xiao}
\author[mg]{Ming Feng Gu}
\author[fd]{Chong Yang Chen\corref{cyc}}

\cortext[kw]{wang$_{-}$kai10@fudan.edu.cn}
\cortext[rs]{rsi@fudan.edu.cn}
\cortext[cyc]{chychen@fudan.edu.cn}

\address[fd]{Shanghai EBIT Lab, Key Laboratory of Nuclear Physics and Ion-beam Application, Institute of Modern Physics, Department of Nuclear Science and Technology, Fudan University, Shanghai 200433, China}
\address[hb]{Hebei Key Lab of Optic-electronic Information and Materials, The College of Physics Science and Technology, Hebei University, Baoding 071002, China}
\address[bls]{Spectroscopy, Quantum Chemistry and Atmospheric Remote Sensing (SQUARES), CP160/09, Universit\'{e} libre de Bruxelles, 1050 Brussels, Belgium}
\address[rd]{Department of Materials Science and Applied Mathematics, Malm{\"o} University, SE-20506, Malm{\"o}, Sweden}
\address[mg]{Space Science Laboratory, University of California, Berkeley, CA 94720, USA}

\begin{abstract}
Atomic properties of $n=3$ states of the W$^{56+}$ $-$ W$^{61+}$ ions are systematically investigated through two state-of-the-art methods, namely, the second-order many-body perturbation theory, and the multi-configuration Dirac-Hartree-Fock method combined with the relativistic configuration interaction approach. The contributions of valence-valence and core-valence electron correlations, the Breit interaction, the higher-order retardation correction beyond the Breit interaction through the transverse photon interaction, and the quantum electrodynamical corrections to the excitation energies are studied in detail.
The excitation energies and wavelengths obtained with the two methods agree with each other within \SI{\approx0.01}{\%}. The present results achieve spectroscopic accuracy and provide a benchmark test for various applications and other theoretical calculations of W$^{56+}$ $-$ W$^{61+}$ ions. They will assist spectroscopists in their assignment and direct identification of observed lines in complex spectra.
\end{abstract}


\end{frontmatter}

\twocolumn
\section{Introduction}

Motivated by the potential use in plasma diagnostics in the future tokamak fusion reactor ITER, $M$-shell W$^{56+}$ $-$ W$^{61+}$ tungsten ions are at present subject of extensive research
~\citep{Peacock.2008.V86.p277,Skinner.2009.VT134.p14022,Beiersdorfer.2010.V43.p144008,
Beiersdorfer.2015.V3.p260,Rzadkiewicz.2018.V97.p52501,Zhao.2018.V119.p314,Safronova.2018.V97.p12502}.  Tungsten ions can be transported from the relatively cold divertor region to the plasma core with temperatures on the order of \SI{20}{keV}, and ionize to $M$-shell charge states, which are expected to strongly emit in the x-ray and extreme ultraviolet (EUV) spectral ranges. Of special interest are the many strong emissions
between states of the $3s^23p^q$ ($q=1-6$) ground configurations in the region of \SI{10}{\AA}~and \SI{180}{\AA}~\cite{Clementson.2010.V81.p52509,Lennartsson.2013.V87.p62505}. The measured radiation can be reliably used to diagnose plasma properties such as temperature and electron density.

There are a few measurements of $n=3 \rightarrow n= 3$ transitions in W$^{56+}$ $-$ W$^{61+}$ ions. Using the electron beam ion trap (EBIT) facility of the National Institute of~Standards and Technology (NIST), Ralchenko $et\ al.$~\cite{Ralchenko.2008.V41.p21003} measured the EUV spectra involving these transitions and ions
in the range of \SI{45}{\AA} -- \SI{180}{\AA}.
Using the EBIT facility of the Lawrence Livermore National Laboratory, Clementson $et\ al.$~\cite{Clementson.2010.V81.p52509} measured accurate wavelengths in the range of \SI{19}{\AA} -- \SI{25}{\AA}.
A further study of the EUV spectra between \SI{27}{\AA} and \SI{41}{\AA} from these highly charged ions (HCIs) of tungsten was reported by Lennartsson $et\ al.$ using the same facility~\cite{Lennartsson.2013.V87.p62505}.
In this experiment, three lines at $\lambda$ = \SI{35.644(4)}{\AA}, \SI{35.668(4)}{\AA} and \SI{34.779(4)}{\AA} were not identified due to the fact that the relativistic configuration-interaction (RCI) calculations used in~\cite{Lennartsson.2013.V87.p62505}, and performed with the Flexible Atomic Code (FAC), were not accurate enough to help the direct line identification.

Given this background, we carried out a high-precision benchmark study of sulfur-like tungsten (W$^{58+}$)~\cite{Zhang.2020.V101.p32509} by using a state-of-the-art method, namely, the multi-configuration Dirac-Hartree-Fock (MCDHF) method combined with the RCI approach~\citep{Fischer.2016.V49.p182004}. The various contributions to excitation energies, such as electron correlations, along with the Breit and transverse photon interactions, were investigated in detail.
We compared the performance of three different quantum electrodynamic (QED) potentials~\cite{Si.2018.V98.p12504,Li.2018.V98.p20502} to estimate the accuracy of QED calculations, and reported improved theoretical excitation energies and transition wavelengths for W$^{58+}$ by at least one order of magnitude.
The accuracy of the MCDHF-RCI calculations~\cite{Zhang.2020.V101.p32509} allowed the authors to assign the unidentified line at $\lambda$~= \SI{34.779(4)}{\AA} from ~\cite{Lennartsson.2013.V87.p62505} to
the M1 transition $3s^23p^4\ ^1D_2$ $\rightarrow$ $3s^23p^4\ ^3P_2$ of S-like W. The present
study is an extension of our recent work on W$^{58+}$~\cite{Zhang.2020.V101.p32509}. We use two methods, MCDHF-RCI and relativistic many-body perturbation theory (RMBPT), to study systematically the properties of tungsten with higher degree of ionization, up to W$^{61+}$.
Using these two methods (MCDHF-RCI and RMBPT), two sets of theoretical
excitation energies achieving spectroscopic accuracy are reported for W$^{56+}$ $-$ W$^{61+}$. The present two data sets agree with each other within \SI{\approx0.01}{\%}.

In the MCDHF calculations the wave functions are expanded in
terms of  configuration state functions (CSFs) generated by orbital substitutions
from configurations in the mulireference space, to a set of orbitals.
By increasing systematically the latter, the effect of valence-valence (VV) and core-valence (CV) electron correlations on the considered properties can be monitored.
What makes MCDHF calculations challenging is the rapid increase in the number of CSFs as a function of the increasing active set and the number of correlated electrons.

To the contrary, electron correlations, including CV and core-core (CC), can be effectively captured through second-order in the RMBPT calculations based on the relativistic many-body perturbation theory~\cite{Lindgren.2010.V108.p2853}.

In the present MCDHF calculations, the Breit and transverse photon interactions, and QED corrections are included, step by step, to study in detail their contributions to the excitation energies.
In addition, some suggestions of assignments are made for the other two unidentified lines at $\lambda$ = \SI{35.644(4)}{\AA} and \SI{35.668(4)}{\AA} from~\cite{Lennartsson.2013.V87.p62505}.
This work paves the way for future applications targeting accurate theoretical prediction of properties of a very wide range of HCIs, and  provides precision benchmarks for spectra identification.

As an introduction to our work, we resume the previous relevant theoretical studies. Using the general-purpose relativistic atomic structure package (GRASP89)~\cite{Dyall.1989.V55.p425} and the flexible atomic code (FAC)~\cite{Gu.2008.V86.p675},
Aggarwal and Keenan~\cite{Aggarwal.2014.V100.p1603,Aggarwal.2016.V111-112.p187} calculated excitation energies, radiative rates, and lifetimes for the W$^{57+}$ $-$ W$^{61+}$ ions.
Using the GRASP89 code~\cite{Dyall.1989.V55.p425}, Xu $et\ al.$~\cite{Xu.2017.V95.p283} reported excitation energies, wavelengths and transition probabilities for the same ions. While the above calculations~\cite{Aggarwal.2014.V100.p1603,Aggarwal.2016.V111-112.p187,Xu.2017.V95.p283} tend to support line identifications, none of them are high-precision calculations. For example, the excitation energies for W$^{58+}$ from ~\cite{Xu.2017.V95.p283} differ by
3000 -- \SI{70000}{cm}$^{-1}$ from the results of~\cite{Aggarwal.2016.V111-112.p187}.
	
15 wavelengths of the $n = 3 - 3$ transitions for Ne- to Ar-like tungsten were provided by Chen and Cheng~\cite{Chen.2011.V84.p12513} using the RCI method. Their calculation is based on the relativistic no-pair Hamiltonian and uses finite B-spline orbitals in a cavity as basis functions. Safronova $et\ al.$~\cite{Safronova.2010.V43.p74026} and Ekman $et\ al.$~\cite{Ekman.2018.V120.p152} calculated excitation energies and transition probabilities for W$^{61+}$ using the RMBPT and MCDHF-RCI methods, respectively.
Using the GRASP0 code~\cite{Grant.1980.V21.p207}, excitation energies and transition probabilities for W$^{57+}$  were estimated by Mohan $et\ al.$~\cite{Mohan.2014.V92.p177}.
The energy of $3p_{3/2} - 3p_{1/2}$ fine splitting in W$^{57+}$ was studied by Kozio{\l}~\cite{Koziol.2020.V242.p106772} using GRASP2K~\cite{Joensson.2013.V184.p2197}, MCDFGME~\cite{Gorceix.1987.V20.p639,Indelicato.1987.V20.p651} and FAC~\cite{Gu.2008.V86.p675} codes.

The paper is organized as follows.
The MCDHF-RCI and RMBPT methods are outlined in Sec.~\ref{Outline of theory}. In Sec.~\ref{results and discussion} we present our numerical results and compare them with measured values and previous calculations. Sec.~\ref{conclusions} is a brief summary.

If not explicitly indicated, atomic units are used throughout this work.

\section{Outline of theory}~\label{Outline of theory}
\subsection{MCDHF}~\label{MCDHF}
\subsubsection{Electron correlation}~\label{EC}

In the MCDHF method~\citep{Grant.2007.V.p}, electron correlation is included by expanding the wave function $\Psi\left(\Gamma J \pi \right)$ in CSFs
\begin{equation}
	\Psi\left(\Gamma J \pi \right)=\sum_{i=1}^Mc_i\Phi\left(\gamma_i  J \pi \right).
\end{equation}
where, for each CSF, $\Phi\left(\gamma_i J \pi \right)$, $\gamma_i$ specifies the occupied subshells with their complete angular coupling tree information, $J$ and $\pi$ being, respectively, the total angular momentum and the parity. Each CSF is a $jj$-coupled  many-electron function built from antisymmetrized products of one-electron Dirac orbitals~\citep{Grant.2007.V.p}.
The large and small components of the radial parts of one-electron orbitals and the expansion coefficients \{$ c_i $\} of the CSFs are obtained by solving iteratively the Dirac-Hartree-Fock radial equations and the configuration interaction eigenvalue problem resulting from the application of the variational principle on the energy functional of the targeted states in the extended optimal level (EOL) scheme~\cite{Grant.2007.V.p,Grant.1980.V21.p207}.
The energy functional is based on the Dirac-Coulomb (DC) Hamiltonian
\begin{eqnarray}
\label{eq:DC}
\hspace{-2.0cm}
H_{\mathrm{DC}}
& = &
\sum_{i=1}^N \left( c\ {\bm{\alpha}}_i \cdot {\bf{p}}_i
+ V_{nuc}(r_i) + c^2 (\beta_i-1)  \right)   \nonumber \\
&+& \!\sum_{j>i=1}^N \frac{1}{r_{ij}},
\end{eqnarray}
and accounts for relativistic kinematic effects.

For Al-like W$^{61+}$, Si-like W$^{60+}$ and P-like W$^{59+}$, all possible $3s^u 3p^v 3d^w$ configurations with $k=u+v+w=$~3, ~4~and 5, respectively, constitute the MR spaces.
For S-like W$^{58+}$, Cl-like W$^{57+}$ and Ar-like W$^{56+}$, the MR spaces include the following configurations
\{$3s^23p^{k-2}$, $3s^23p^{k-3}3d$, $3s^23p^{k-4}3d^2$, $3s^23p^{k-5}3d^3$, $3s3p^{k-2}3d$, $3s3p^{k-3}3d^2$, $3s3p^{k-4}3d^3$\} with $k =$~6,~7~and 8, respectively.
Additional configurations that differ from one ion to another complete these MR spaces:
\{$3p^43d^2$, $3p^53d$, $3p^6$\} for S-like W$^{58+}$, \{$3p^43d^3$, $3s3p^6$, $3s3p^23d^4$\} for Cl-like W$^{57+}$, and \{$3s^23p^23d^4$\} for Ar-like W$^{56+}$.
The CSF expansions are generated by allowing single (S) and double (D) excitations of all the $n = 3$  electrons, namely valence electrons, from all MR configurations to
orbitals with $n\leq7, l\leq5$ ($i.\ e.$, up to $h$-orbital symmetry). These CSFs describe the VV electron correlation.
In a second series of calculations we add, to these expansions, CSFs resulting from  SD-MR substitutions of all $n = 2,~3$ electrons to
orbitals with $n \leq 6$, $l \leq 5$, with the restriction of allowing maximum one hole in the $n=2$ core shell. These added CSFs describe the CV correlation effects.
For the present calculations of the W$^{56+}$ $-$ W$^{61+}$ ions, the numbers of CSFs distributed over different $J^{\pi}$ symmetries in the final state expansions are between \SI{5838070} and \SI{46392382} for even parity, and between \SI{6568665} and \SI{2215477} for odd parity.
The largest number of CSFs distributed over one $J^{\pi}$ symmetry is up to \SI{9358525}(for $3^+$ symmetry of W$^{56+}$). To solve effectively the eigenpairs for such large CSFs expansion, some modifications are made for the original GRASP2K code~\cite{Joensson.2013.V184.p2197}. For example, the BLAS subroutines ddot and dgemv, which are employed calculating respectively the vector-vector dot product and matrix-vector multiplication, are parallelized by using message passing interface (MPI).
To speed up the Gram-Schmidt re-orthogonalization process within the matrix diagonalization, the DVDSON routine is modified to permit that the targeted eigenpairs can be searched one by one.
These modifications can save a lot of time especially for a very large CSFs expansion.
	
By comparing the present MCDHF results with our RMBPT calculations (see below, Section~\ref{RMBPT})
that consider additionally core-core electron excitations,
we infer that the latter are unimportant for the excitation energies of the studied states and are therefore neglected in the MCDHF calculations. No substitutions are allowed from the $1s$ shell, which defines an inactive closed $1s^2$ core.
Convergence of MCDHF excitation energies, monitored by enlarging the active set in a systematical way~\cite{Guo.2016.V93.p12513}, is achieved within \SI{0.01}{\%}.

\subsubsection{Breit and QED Corrections}~\label{Breit}
	In the relativistic description of the many-electron system, the Dirac-Coulomb Hamiltonian is the starting point that should be corrected by the so called transverse photon (TP) interaction, which, in the $\alpha^2$ approximation, takes the form:
    \begin{eqnarray}
\label{eq:TP}
H_{\mathrm{TP}} & = & -\mathop{\sum}\limits ^N \limits_{i<j}
\left[ \frac{\boldsymbol{ \alpha}_{i} \cdot \boldsymbol{\alpha}_{j}}  {r_{ij}}  \cos(\omega_{ij}r_{ij}/c) \right. \nonumber \\
& - & \left.  (\boldsymbol{\alpha}_{i} \cdot \boldsymbol{\nabla}_{i})
 (\boldsymbol{\alpha}_{j} \cdot \boldsymbol{\nabla}_{j})\frac{\cos(\omega_{ij}r_{ij}/c)-1}{\omega^2_{ij}r_{ij}/c^2}
 \right],
\end{eqnarray}
where $\omega_{ij}$ is the frequency of the exchanged virtual photon propagating the interaction.

In the low-frequency limit $\omega_{ij}\rightarrow 0$, the TP interaction reduces to the Breit interaction~\cite{Grant.1976.V9.p761}
\begin{equation}
\label{eq:Breit}
H_{\mathrm{Breit}}
  =  - \sum_{j>i=1}^N \; \frac{1}{2 r_{ij}} \left[ \left(  \bm{\alpha}_i \cdot \bm{\alpha}_j \right) +
\frac{\left( {\bm{\alpha}}_i \cdot {\bf r}_{ij} \right)\left( \bm{\alpha}_j \cdot {\bf r}_{ij} \right)}{r^2_{ij}}\right],
\end{equation}
which is the sum of the Gaunt interaction and the Breit retardation~\cite{Indelicato.2007.V45.p155}.

The higher-order (HO) retardation correction beyond the Breit interaction (\ref{eq:Breit}) is therefore defined as the difference
\begin{equation}
\label{eq:HO_retardation}
H^{\mathrm{HO}}
\equiv
H_{\mathrm{TP}} - H_{\mathrm{Breit}}.
\end{equation}

Once the orbitals optimized through the MCDHF procedure are available, the Breit interaction, higher-order retardation  correction, and the leading QED effects (self-energy and vacuum polarization) can be added to the Dirac-Coulomb Hamiltonian in the RCI calculations to capture relativistic corrections to the Coulomb interaction.

The evaluation of matrix elements of the TP Hamiltonian is non-trivial.
Following the prescription in ~\citep{Grant.2007.V.p}, the matrix elements
are obtained by averaging terms involving frequencies $\omega_{ij}$ computed
from differences in left-hand and right-hand side one-electron orbital energies. However, the orbital energies are physically meaningful only for spectroscopic orbitals. For this reason,
in the present work, the TP Hamiltonian matrix elements
include the frequency-dependent contributions when involving only the spectroscopic orbitals
\[ \{ 1s_{1/2},~2s_{1/2}, ~2p_{1/2}, ~2p_{3/2}, ~3s_{1/2}, ~3p_{1/2}, ~3p_{3/2}, ~3d_{3/2}, ~3d_{5/2} \} \]
spanning the MR configurations. For contributions involving any of the so-called correlation orbitals that are unoccupied in the MR subspace, but appear in the orbital sets for describing electron correlation excitations,
the low frequency limit $\omega_{ij}\rightarrow 0$ is considered.

The current status of bound state QED calculations of transition energies for few-electron highly-charged ions has recently been reviewed by Indelicato~\cite{Indelicato.2019.V52.p232001}.
The one-loop QED correction is separated into two components, namely, the VP and the SE. The one-electron VP effect can be represented by a local potential. The analytical expressions derived by Fullerton and Rinker~\cite{Fullerton.1976.V13.p1283} are used in this work for the Uehling model potential and the higher-order K$\ddot{\rm{a}}$ll$\acute{\rm{e}}$n-Sabry VP potential. The Wichmann-Kroll (WK) potential is estimated with the help of the approximate formulas derived in~\cite{Fainshtein.1991.V24.p559}.
For W$^{56+}$ $-$ W$^{61+}$, the SE contribution dominates the QED corrections. Three different models (M1 - M3) are investigated for estimating the self-energy contribution:

\begin{itemize}
\item

SE - M1: In the GRASP2K code~\cite{Joensson.2013.V184.p2197}, the
one-electron SE contribution of an orbital $\psi$ is estimated from
\begin{equation}\label{SE}
E_{\rm{SE}}=S_{\rm{M1}}E^{\rm{H}}_{\rm{SE}},
\end{equation}
where
\begin{equation}\label{SE_form}
E^{\rm{H}}_{\rm{SE}} =  \frac{\alpha}{\pi}  (Z \alpha)^4 F (Z \alpha) m_e c^2 \; ,
\end{equation}
is the hydrogenic SE value calculated by Mohr and
collaborators~\cite{Mohr.1974.V88.p26,Mohr.1983.V29.p453,Mohr.1992.V45.p2727},
and where the screening factor $S_{\rm{M1}}$ is determined by the ratio of the mean radius of each MCDHF orbital in a small region around the nucleus $r < \SI{0.0219}{a_0}$ to the equivalent mean radius for a hydrogenic orbital
\begin{equation}\label{SE}
S_{\rm{M1}}=\cfrac{\langle\psi_{r<0.0219}|r |\psi_{r<0.0219}\rangle}{\langle\psi_{r<0.0219}^{\rm{H}}|r
|\psi_{r<0.0219}^{\rm{H}}\rangle} \; .
\end{equation}
The total SE contribution is given as a sum of one-electron corrections weighted by the fractional occupation number of the one-electron orbital $\psi$ in the wave function.

\item

SE - M2: A self-energy screening approximation, based on the Welton interpretation~\cite{Welton.1948.V74.p1157} has been implemented in GRASP2K~\cite{Joensson.2013.V184.p2197} by Lowe {\it et al.}~\cite{Lowe.2013.V85.p118}. In this approach, the latest available hydrogenic values~\cite{Mohr.1992.V45.p2727,LeBigot.2001.V64.p52508}, modified to account for finite-nuclear-size effects~\cite{Mohr.1993.V70.p158,Beier.1998.V58.p954} are used to evaluate the SE contribution to a one-electron orbital $\psi$ from
\begin{equation}\label{SE}
E_{\rm{SE}}=S_{\rm{M2}}E^{\rm{H}}_{\rm{SE}},
\end{equation}
where now the screening factor $S_{\rm{M2}}$ is determined
by comparing the mean $\nabla^2V(\boldsymbol{x})$ of each MCDHF
orbital to that of a hydrogenic orbital

\begin{equation}
\label{Welton}\\
S_{\rm{M2}}=\cfrac{\langle\psi|\nabla^2V(\boldsymbol{x})|\psi\rangle}{\langle\psi^{\mathrm{H}}|\nabla^2V(\boldsymbol{x})
|\psi^{\mathrm{H}}\rangle}.\\
\end{equation}
This corresponds to option (A.2) of \cite{Lowe.2013.V85.p118}.

\item

SE - M3: A model QED operator, which also includes the non-local QED part to evaluate the SE corrections for many-electron atomic systems, has been developed by Shabaev $et\  al.$~\cite{Shabaev.2013.V88.p12513,Shabaev.2015.V189.p175}. It was implemented in GRASP2K to study the  ground-term fine structures of F-like~\cite{Li.2018.V98.p20502}
and Co-like ions~\cite{Si.2018.V98.p12504}
and used in our recent benchmark study of S-like tungsten~\cite{Zhang.2020.V101.p32509}.
The SE correction is determined by evaluating the matrix element of the model self-energy operator
\begin{equation}
\begin{array}{ll}\label{Shabaev}\\
h^{\mathrm{SE}}=
h^{\mathrm{SE}}_{\mathrm{loc}}
+\mathop{\sum}\limits ^n \limits_{i,j=1} |\phi_i  \rangle \bigg\lbrace
\mathop{\sum}\limits ^n \limits_{k,l=1} [(S^t)^{-1}]_{ik} \\
\times \langle \psi_k| \big\lbrace \frac{1}{2}
 [ \sum (\varepsilon_k) +\sum(\varepsilon_l) ]
-h^{\mathrm{SE}}_{\mathrm{loc}} \big\rbrace
 | \psi_l \rangle
 (S^{-1})_{lj} \bigg\rbrace
 \langle\phi_j|
\end{array}
\end{equation}
with the many-electron wave function.
$ h^{\mathrm{SE}}_{\mathrm{loc}}$
is a quasi-local operator acting differently on wave functions of different angular symmetry, $\{\psi_i(\mathbf{r})\}^n_{i=1}$ is the basis of the hydrogenic wave functions, $\{\phi_i(\mathbf{r})\}^n_{i=1}$ is the model basis of projected functions,
and $S_{ik}$ are the overlap matrix elements $S_{ik} = \langle \phi_i | \psi_k \rangle$ (see original work~\cite{Shabaev.2013.V88.p12513} for more details).
\end{itemize}

The notations of the various correlation models are the same as in our recent work~\cite{Zhang.2020.V101.p32509}.	The MCDHF calculations based on the Dirac-Coulomb Hamiltonian are labeled VV when only including VV electron correlation, and +CV when further considering CV electron correlation.
The DCB notation is used for Dirac-Coulomb Breit results obtained by adding in RCI the Breit interaction (\ref{eq:Breit}) to the Dirac-Coulomb Hamiltonian using the VV+CV CSF expansions and the MCDHF orbitals.
Calculations beyond the Breit interaction, $i.\ e.$ including higher-order retardation correction by adding the frequency-dependent transverse photon interaction (\ref{eq:TP}) to the DC  Hamiltonian, are labeled DCTP.
The calculations are labeled +QED when including QED corrections to ${H}_{\mathrm{DCTP}}$ in the very last step.

\subsection{RMBPT}~\label{RMBPT}
The RMBPT method~\cite{Lindgren.1974.V7.p2441,Safronova.1996.V53.p4036,Vilkas.1999.V60.p2808} implemented in the FAC code~\cite{Gu.2008.V86.p675}
has been successfully used to calculate atomic data of high accuracy in our recent studies~\cite{Guo.2016.V93.p12513,Si.2016.V227.p16,Si.2018.V239.p3,Zhang.2018.V121-122.p256,Zhang.2018.V206.p180,Li.2019.V126.p158,Li.2020.V133-134.p101339,Wang.2014.V215.p26,Wang.2015.V218.p16,Wang.2018.V239.p30,Wang.2018.V235.p27,Wang.2020.V246.p1}.  In this method, a limited number of configurations can be included in the RCI expansion.
The complete positive energy state space of the $H_{\rm{DCB}}$ is then split
into a model Hamiltonian $H_0$ and a perturbation $ H'$, $i.\ e.$, $H_{\rm{DCB}}= H_0+H'$.
The eigenfunctions $\Phi_k^0$ and eigenvalues $E_k^0$ of $H_0$ can be calculated using
\begin{equation}
H_0\Phi_k^0=E_k^0\Phi_k^0.
\end{equation}
The eigenfunctions $\Phi_k^0$ of $H_0$ are divided into a model space, $M$, and the orthogonal space, $N$.
An ``effective Hamiltonian" $H_{\rm{eff}}$, in the $M$ space, the eigenvalues of
which are the true eigenenergies of the full Hamiltonian, is defined as
\begin{equation}
	H_{\rm{eff}}=PH_0P+PH'\Omega.
\end{equation}
Here $P$ is a projection operator that produces a state $\Psi_k^0$ in the $M$ space when it operates on an eigenfunction $\Psi_k$ of the full Hamiltonian, where $\Psi_k^0$ is a linear combination of the subset of $\Phi_k^0$ that belongs to the $M$ space.
$\Omega$ is the wave operator that transforms $\Psi_k^0$ back to $\Psi_k$.

The matrix element of the first-order effective Hamiltonian $H_{\rm{eff}}$ is
\begin{equation}
\langle \Phi\left(\gamma_i  J \pi \right) | H^{(1)}_{\rm{eff}}| \Phi\left(\gamma_j  J \pi
\right) \rangle= (H_{\rm{DCB}})_{ij} + \sum_{r\in N}\frac{H'_{ir} H'_{rj}}{E^0_{j}-E^0_{r}},
\end{equation}
here
$(H_{\rm{DCB}})_{ij} =\langle \Phi\left(\gamma_i  J \pi \right) | H_{\rm{DCB}} | \Phi\left(\gamma_j  J \pi \right) \rangle$, $H'_{ir}=\langle \Phi\left(\gamma_i  J \pi \right) | H' | \Phi\left(\gamma_r  J \pi \right) \rangle$ ($i, j\in M, r\in N$).
By solving the generalized eigenvalue problem for the first-order effective Hamiltonian $H_{\rm{eff}}$, we can obtain the eigenvalues in second order.
	
The Breit interaction, the higher-order retardation correction, and the leading QED corrections, with SE implemented according to model M3, are also considered in the same way as for the MCDHF calculations.

In the present RMBPT calculations, all possible $3s^u 3p^v 3d^w$ configurations with $k=u+v+w=3~-~8$ are contained in the $M$ space for W$^{56+}$ $-$ W$^{61+}$.
Configurations that are generated from SD excitations of the $M$ space are involved in the $N$ space. For S and D excitations the maximum $n$ values are 125 and 65 respectively, and the maximum $l$ value is 20.
Besides VV and CV electron correlation included in the MCDHF calculations, the CC electron correlation effects are further considered in the RMBPT calculations, and substitutions are also allowed from the $1s$ shell.

It should be noted that the wave functions for the levels are given as expansions over $jj$-coupling CSFs in both of the MCDHF-RCI and RMBPT calculations. In order to match the calculated states against the NIST ASD and other theoretical calculations, the wave functions are transformed from a $jj$-coupling CSF basis into a $LSJ$-coupling CSF basis using the methods developed by Gaigalas $et\ al.$~\cite{Gaigalas.2004.V157.p239,Gaigalas.2017.V5.p6}.

\section{Results and discussions}~\label{results and discussion}

\subsection{Excitation energies}~\label{transition}
\subsubsection{Electron correlation}

17 soft x-ray spectral lines of the $3p_{1/2} - 3d_{3/2}$ and $3s_{1/2} - 3p_{3/2}$ transitions between \SI{19}{\AA} and \SI{25}{\AA} were measured from  W$^{56+}$ to W$^{61+}$ by Clementson $et\ al.$~\cite{Clementson.2010.V81.p52509}.
The original line labels consisting in an isoelectronic identifier followed by an integer (Al-1, Al-2, etc.) are shown in Table~\ref{configuration}.
All these transitions are connected with the ground states of different tungsten ions.
All these spectral lines belong to E1 transitions that connect with the ground states of different tungsten ions.
Using these measured wavelengths, experimental excitation energies $E_{\rm expt}$ for 17 levels are calculated. These levels formed by exciting one electron $3p_{1/2}$ to the $3d_{3/2}$ subshell belong to the $3s^23p^v3d~(v~=~0~-~5)$ configurations. The other levels are formed by exciting one electron $3s_{1/2}$ to the $3p_{3/2}$ subshell, and belong to the $3s3p^v~(v~=~2~-~6)$ configurations.

\begin{figure}[htb!]		
\centering
\includegraphics[width=6.0in]{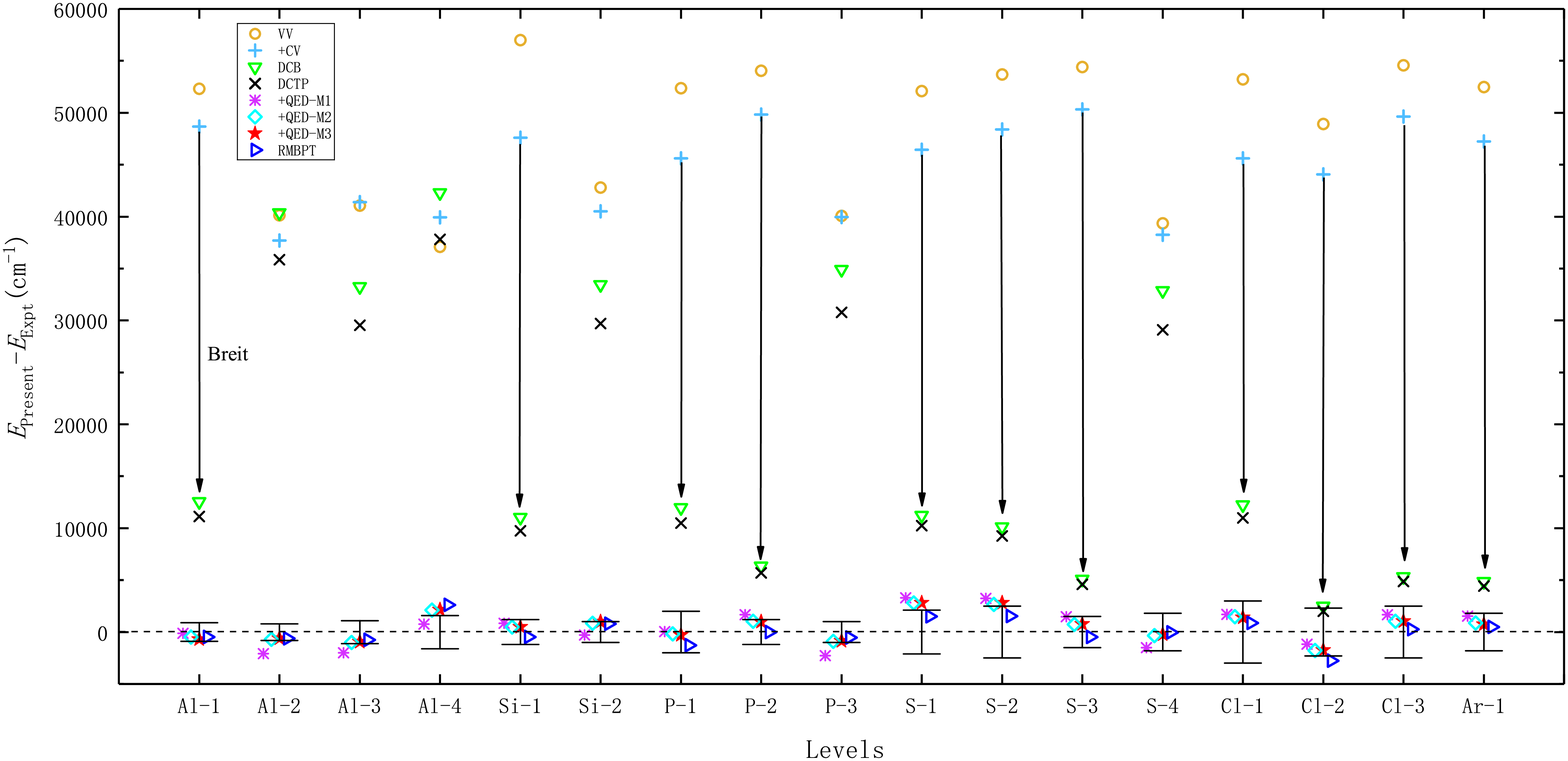}	
\caption{\label{PRA2010_E_effects}The deviations $\Delta E$ (in cm$^{-1}$) of the present MCDHF-RCI excitation energies from experimental values~\cite{Clementson.2010.V81.p52509}. The meaning of the labels VV, +CV, DCB, DCTP +QED is same as in Table~\ref{Table3pk_E-NIST}. The differences between the final results from our RMBPT calculations and experimental values are also shown. The lengths of vertical arrows depict the contribution of Breit interactions to excitation energies of the $3s^23p^v3d~(v~=~0~-~5)$ levels. The levels without vertical arrows belong to the $3s3p^v~(v~=~2~-~6)$ configurations. All the corresponding data are available in Table~\ref{Table3pk_E-NIST}. Vertical error bars are experimental uncertainties.}
	\end{figure}

Experimental values and excitation energies from the above correlation and interaction models of our MCDHF-RCI calculations for these 17 levels, as well as the final results from our RMBPT calculations, are listed in Table~\ref{Table3pk_E-NIST}.
The deviations ($\Delta E_{\rm{MCDHF-RCI}}=E_{\rm{MCDHF-RCI}}-E_{\rm{Expt}}$) between our calculated {MCDHF-RCI} excitation energies and the experimental values, together with the deviations ($\Delta E_{\rm{RMBPT}}=E_{\rm{RMBPT}}-E_{\rm{Expt}}$), are reported in Table~\ref{Table3pk_E-NIST} as well. To get an overview, these deviations are also shown in Figure~\ref{PRA2010_E_effects}.
To distinguish levels belonging to the $3s^23p^v3d~(v~=~0~-~5)$ configuration from
levels belonging to the $3s3p^v~(v~=~2~-~6)$ configuration, the effect of the Breit interaction for the former are marked by vertical arrows.

Generally, for transitions involving valence excitations, CV electron correlation plays a smaller role than VV electron correlation.
Therefore, CV electron correlation was omitted in previous theoretical calculations, such as Ref.~\cite{Aggarwal.2014.V100.p1603,Aggarwal.2016.V111-112.p187,Xu.2017.V95.p283,Mohan.2014.V92.p177}.
However, in this work we find that contributions of CV electron correlation to excitation energies are grouped according to the electronic configuration of considered levels.
For the $3s3p^v~(v~=~2~-~6)$ levels, as can be seen in Figure~\ref{PRA2010_E_effects} and further
inferred by comparing the two columns $\Delta E_{\rm{MCDHF-RCI}}$ (VV) and $\Delta E_{\rm{MCDHF-RCI}}$ (+CV) of Table~\ref{Table3pk_E-NIST}, CV electron correlation change the excitation energies by $\simeq$~ \SI{100}{cm}$^{-1}$ to \SI{2800}{cm}$^{-1}$.
However, the excitation energies of $3s^23p^v3d~(v~=~0~-~5)$ levels are reduced by \SI{3700}{cm}$^{-1}$ to \SI{9400}{cm}$^{-1}$, and the addition of CV to the VV electron correlation makes the results for these levels closer to observation.
Considering the fact that the estimated uncertainties provided by Kramida~\cite{Kramida.2011.V89.p551}, and included in NIST~\cite{Kramida.2018.V.p}, for the measured excitation energies are as low as several hundreds of cm$^{-1}$, we gather that limiting electron correlation to VV electron correlation is not enough to reach the needed accuracy for assisting spectroscopists in the spectral lines identification process, but CV effects need to be accounted for. This is especially true for the $3s^23p^v3d~(v~=~0~-~5)$ levels formed by exciting one electron $3p_{1/2}$ to the $3d_{3/2}$ subshell. The importance of CV electron correlation is also discussed in our recent work~\cite{Zhang.2020.V101.p32509} and Ref.~\cite{Lennartsson.2013.V87.p62505}.

\subsubsection{The Breit interaction and QED corrections}

Comparing the results in columns
$\Delta E_{\rm{MCDHF-RCI}}$ (+CV) and $\Delta E_{\rm{MCDHF-RCI}}$ (DCB) of Table~\ref{Table3pk_E-NIST}, we can see the contributions from the Breit interaction to the excitation energies, which are also shown in Figure~\ref{PRA2010_E_effects}.
It is found that the contribution of the Breit correction strongly depends on the electronic configuration.
The Breit interaction reduces excitation energies of the $3s^23p^v3d~(v~=~0~-~5)$ levels by
$\simeq$~\SI{33000} - \SI{45000}{cm}$^{-1}$, which further close significantly the gap between the MCDHF-RCI excitation energies of $3s^23p^v3d~(v~=~0~-~5)$ levels and experimental data.
The corresponding effect on the $3s3p^v~(v~=~2~-~6)$ levels is considerably smaller, from \SI{-8000}{cm}$^{-1}$ to \SI{2000}{cm}$^{-1}$.
	
Comparison of the results in columns
$\Delta E_{\rm{MCDHF-RCI}}$ (DCB) and $\Delta E_{\rm{MCDHF-RCI}}$ (DCTP) of Table~\ref{Table3pk_E-NIST} shows that the higher-order frequency-dependent corrections
$H^{\rm{HO}} = H_{\mathrm{TP}} - H_{\rm{Breit}}$
are relatively small compared with the Breit interaction, from \SI{-450}{cm}$^{-1}$ to \SI{-1500}{cm}$^{-1}$, but cannot be neglected for precision calculations, as discussed in our recent work~\cite{Zhang.2020.V101.p32509}.
The ${H}^{\mathrm{HO}}$ contribution to excitation energies is also grouped according to the electronic configuration: the contributions to the $3s3p^v~(v~=~2~-~6)$ levels being larger than those to the $3s^23p^v3d~(v~=~0~-~5)$ levels.
	
When the QED corrections are added to the DCTP calculation, the agreement between our MCDHF-RCI excitation energies and the corresponding observations is improved substantially.
As found in our recent work for sulfur-like tungsten~\cite{Zhang.2020.V101.p32509}, the QED corrections to excitation energies of W$^{56+}$ $-$ W$^{61+}$ are naturally grouped according to the electronic configuration of the level considered.
QED corrections are dominated by contributions from the $s$ electrons, and exciting
the ground configuration $3s$ electron to the $3p$ subshell leads to a huge differential effect ranging from \SI{-29000}{cm}$^{-1}$ to \SI{-36000}{cm}$^{-1}$
for the excitation energies of the $3s3p^v~(v~=~2~-~6)$ levels. The
excitation of the ground state $3p$ electron to the $3d$ subshell leads to a correspondingly smaller effect
ranging from \SI{-4000}{cm}$^{-1}$ to \SI{-12000}{cm}$^{-1}$ for the excitation energies of the $3s^23p^v3d\ (v = 0 - 5)$ levels.

The QED corrections to the excitation energies of W$^{56+}$ $-$ W$^{61+}$ obtained with three different QED methods are reported in Table~\ref{Table3pk_E-NIST} as well. The deviations of the MCDHF-RCI results based on the SE - M1, SE - M2, and SE - M3 from experimental results are also shown in Figure~\ref{PRA2010_E_effects}. As observed in our recent work~\cite{Zhang.2020.V101.p32509}, the values from the SE - M2 and SE - M3 calculations are generally closer to the experimental values than the values from the SE - M1 calculation.
For example, for the $3s3p^2\ ^2P_{1/2}$ level (Al-2) of Table~\ref{Table3pk_E-NIST}, the SE - M2 and SE - M3 results deviate from experiment by \SI{700}{cm}$^{-1}$ (\SI{0.02}{\%}) and
\SI{620}{cm}$^{-1}$ (\SI{0.01}{\%}), respectively (the estimated experimental uncertainties for this level is \SI{800}{cm}$^{-1}$), while the SE - M1 excitation energy deviates from experiment with up to \SI{2070}{cm}$^{-1}$ (\SI{0.05}{\%}).
Generally, the SE - M2 and SE - M3 excitation energies are within or close
to the experimental estimated uncertainties, with average differences $\pm$ standard deviations of \SI{1143}~$\pm$~\SI{753}{cm}$^{-1}$
(\SI{0.024}{\%} $\pm$ \SI{0.016}{\%}) and \SI{1145}~$\pm$~\SI{760}{cm}$^{-1}$ (\SI{0.024}{\%} $\pm$ \SI{0.016}{\%}), respectively. However, the SE - M1 excitation energies differ from the experimental values by \SI{1510}~$\pm$~\SI{910}{cm}$^{-1}$ (\SI{0.032}{\%} $\pm$ \SI{0.019}{\%}).
In the following, the MCDHF-RCI results will be all based on the SE-M3 model which turns to be the best model for describing the self-energy QED correction.

Excitation energies of these 17 levels from the present RMBPT calculations are also reported.
The present RMBPT and MCDHF-RCI results are in excellent agreement with each other on the order of \SI{0.01}{\%}.

It should be noted that for the present results, only the contribution from the diagonal matrix elements of the QED operator is considered.
We indeed investigated the off-diagonal contributions using the FAC code~\cite{Gu.2008.V86.p675} and found the latter to be negligible.

\subsection{Wavelengths}~\label{transition}

Besides the 17 soft x-ray spectral lines between \SI{19}{\AA} and \SI{25}{\AA} for W$^{56+}$ -- W$^{61+}$ measured by Clementson $et\ al.$~\cite{Clementson.2010.V81.p52509},
Lennartsson $et\ al.$~\citep{Lennartsson.2013.V87.p62505} have reported 17 lines of the $3p_{1/2} - 3p_{3/2}$ and $3p_{3/2} - 3d_{5/2}$ transitions in the \SI{33}{\AA} -- \SI{41}{\AA} region, originating from Al-like W$^{61+}$ through Ar-like W$^{47+}$.
In addition, 18 EUV spectra lines of the $3p_{3/2} - 3d_{3/2}$, $3s_{1/2} - 3p_{3/2}$, $3s_{1/2} - 3p_{1/2}$ and $3d_{3/2} - 3d_{5/2}$ transitions for Al-like W$^{61+}$ through Ar-like W$^{56+}$ in the \SI{49}{\AA} -- \SI{174}{\AA} range have been measured by Ralchenko $et~al.$~\cite{Ralchenko.2008.V41.p21003}.
The lines marked with an Greek letter (Al-$\alpha$, Al-$\beta$, etc.) and an English letter (Al-a, Al-b, etc.) are from ~\cite{Lennartsson.2013.V87.p62505} and ~\cite{Ralchenko.2008.V41.p21003}, respectively. The information of all the observed lines from~\cite{Lennartsson.2013.V87.p62505,Ralchenko.2008.V41.p21003} are listed in Table~\ref{configuration}.

Among the previous calculations, Ekman $et\ al.$~\cite{Ekman.2018.V120.p152} and Safronova $et\ al.$~\cite{Safronova.2010.V43.p74026} provide results for Al-like W.
Mohan $et\ al.$~\cite{Mohan.2014.V92.p177} give the lines for Cl-like W and Chen and Cheng~\cite{Chen.2011.V84.p12513} provide the values for three different ions $i.\ e.$, Al-like W, Si-like W, and Ar-like W. The results for Al-like W to Cl-like W ions are reported by Aggarwal and Keenan~\cite{Aggarwal.2014.V100.p1603,Aggarwal.2016.V111-112.p187} and Xu $et\ al.$~\cite{Xu.2017.V95.p283}, respectively.

\begin{figure}[htb!]
\centering
\includegraphics[width=3.8in]{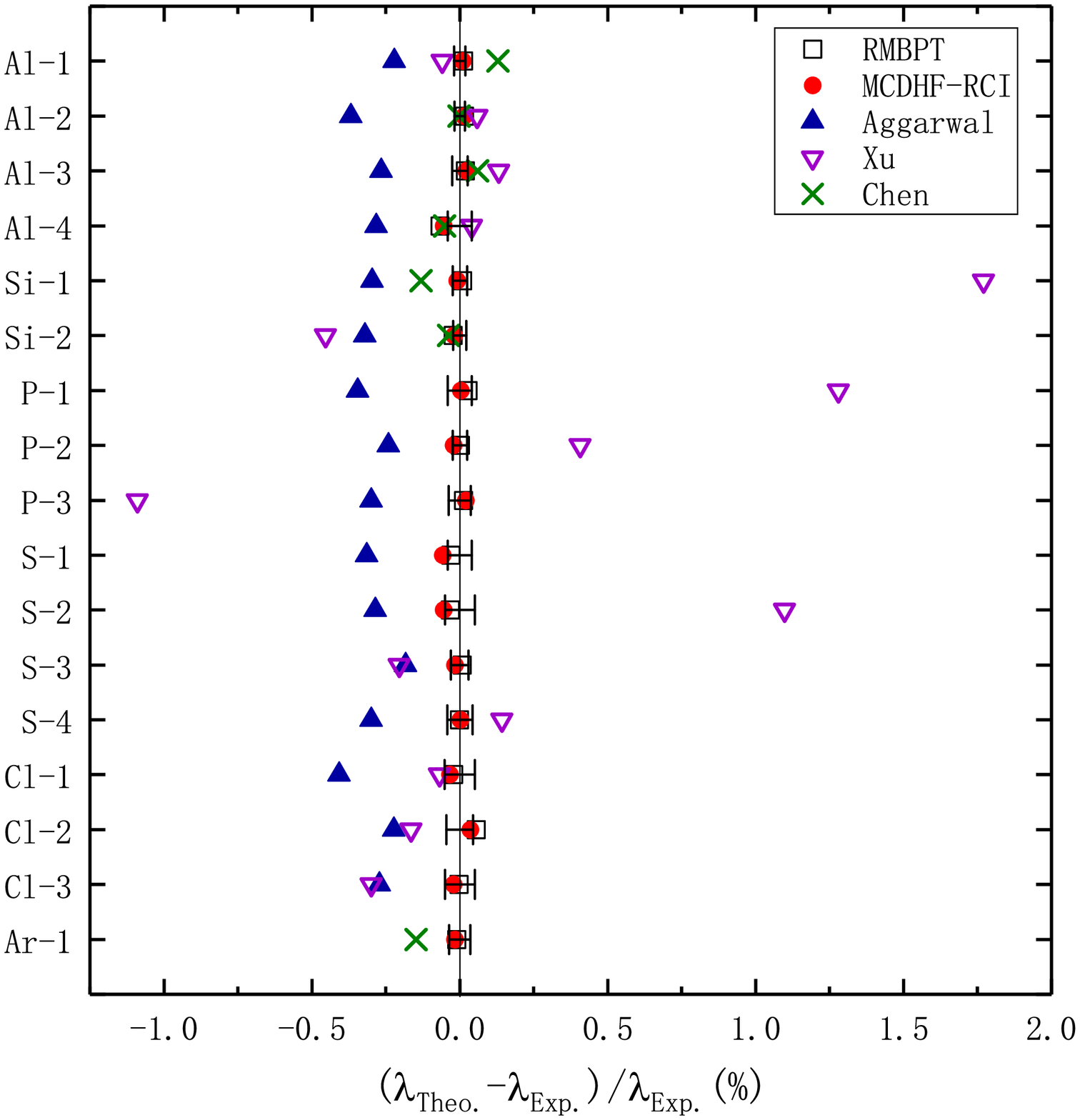}
\caption{\label{fig_PRA2010}The deviations ($\lambda_{\rm{Theo.}}-\lambda_{\rm{Expt.}}$/$\lambda_{\rm{Expt.}}$) in percent of the different theoretical values from the experimental wavelengths in Ref.~\cite{Clementson.2010.V81.p52509} for the $3p_{1/2} - 3d_{3/2}$ and $3s_{1/2} - 3p_{3/2}$ transitions.
Open square  -- the present RMBPT results; solid circle  -- the present MCDHF-RCI results; solid upper triangle -- Aggarwal and Keenan~\cite{Aggarwal.2014.V100.p1603,Aggarwal.2016.V111-112.p187}; open lower triangle -- Xu $et\ al.$~\cite{Xu.2017.V95.p283}; cross -- Chen and Cheng~\cite{Chen.2011.V84.p12513}. Horizontal error bars are experimental uncertainties.}
\end{figure}

\begin{figure}[htb!]
\centering
\includegraphics[width=3.8in]{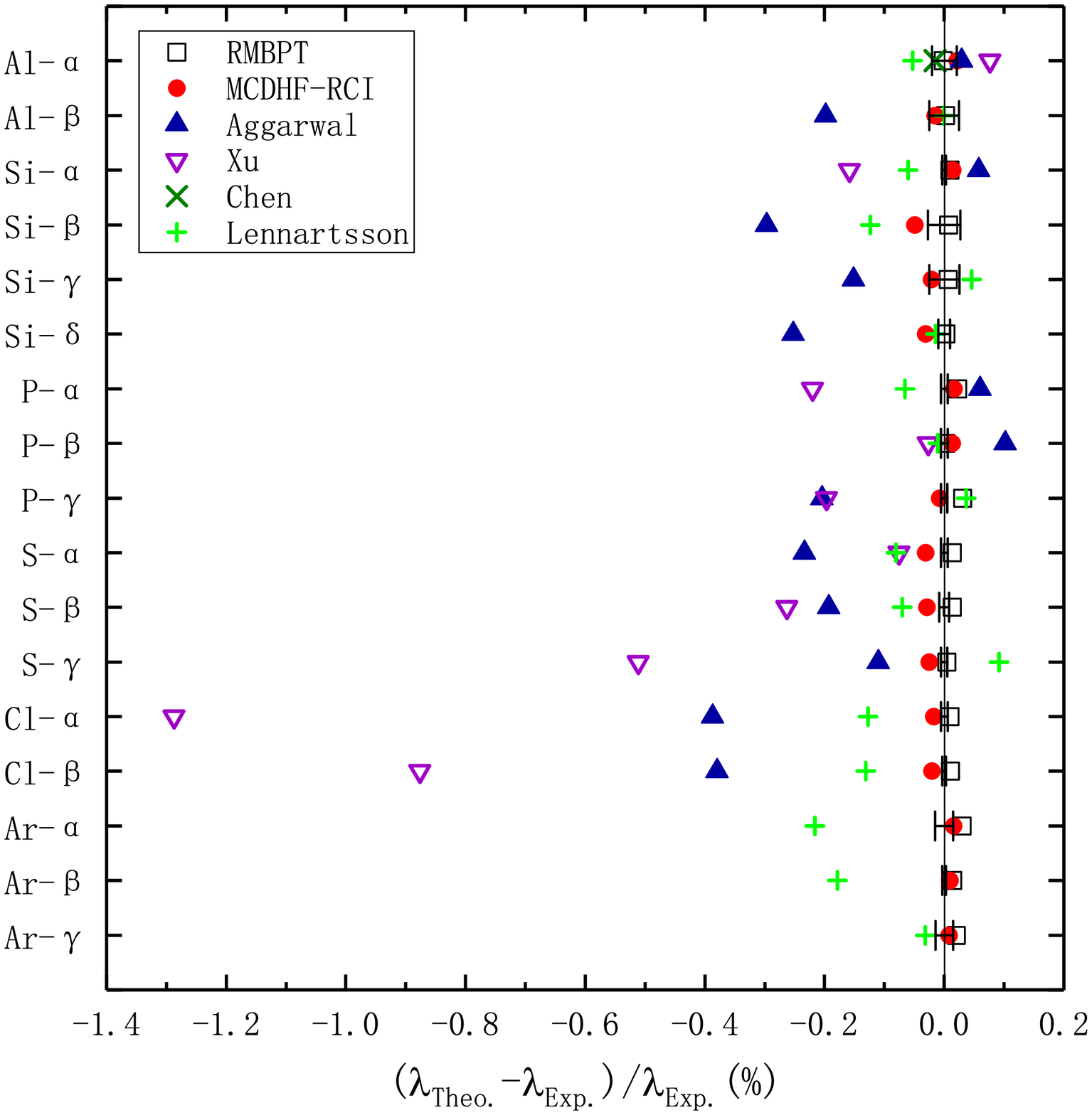}
\caption{\label{fig_PRA2013}The deviations ($\lambda_{\rm{Theo.}}-\lambda_{\rm{Expt.}}$/$\lambda_{\rm{Expt.}}$) in percent of the different theoretical values  from the experimental wavelengths in Ref.~\cite{Lennartsson.2013.V87.p62505} for the $3p_{1/2} - 3p_{3/2}$ and $3p_{3/2} - 3d_{5/2}$ transitions.
Open square  -- the present RMBPT results; solid circle  -- the present MCDHF-RCI results; solid upper triangle -- Aggarwal and Keenan~\cite{Aggarwal.2014.V100.p1603,Aggarwal.2016.V111-112.p187}; open lower triangle -- Xu $et\ al.$~\cite{Xu.2017.V95.p283}; cross -- Chen and Cheng~\cite{Chen.2011.V84.p12513}; plus -- Lennartsson $et\ al.$~\cite{Lennartsson.2013.V87.p62505}. Horizontal error bars are experimental uncertainties. All the corresponding data are available in Table~\ref{Table3pk.tr}.}
\end{figure}

\begin{figure}[htb!]
\centering
\includegraphics[width=3.8in]{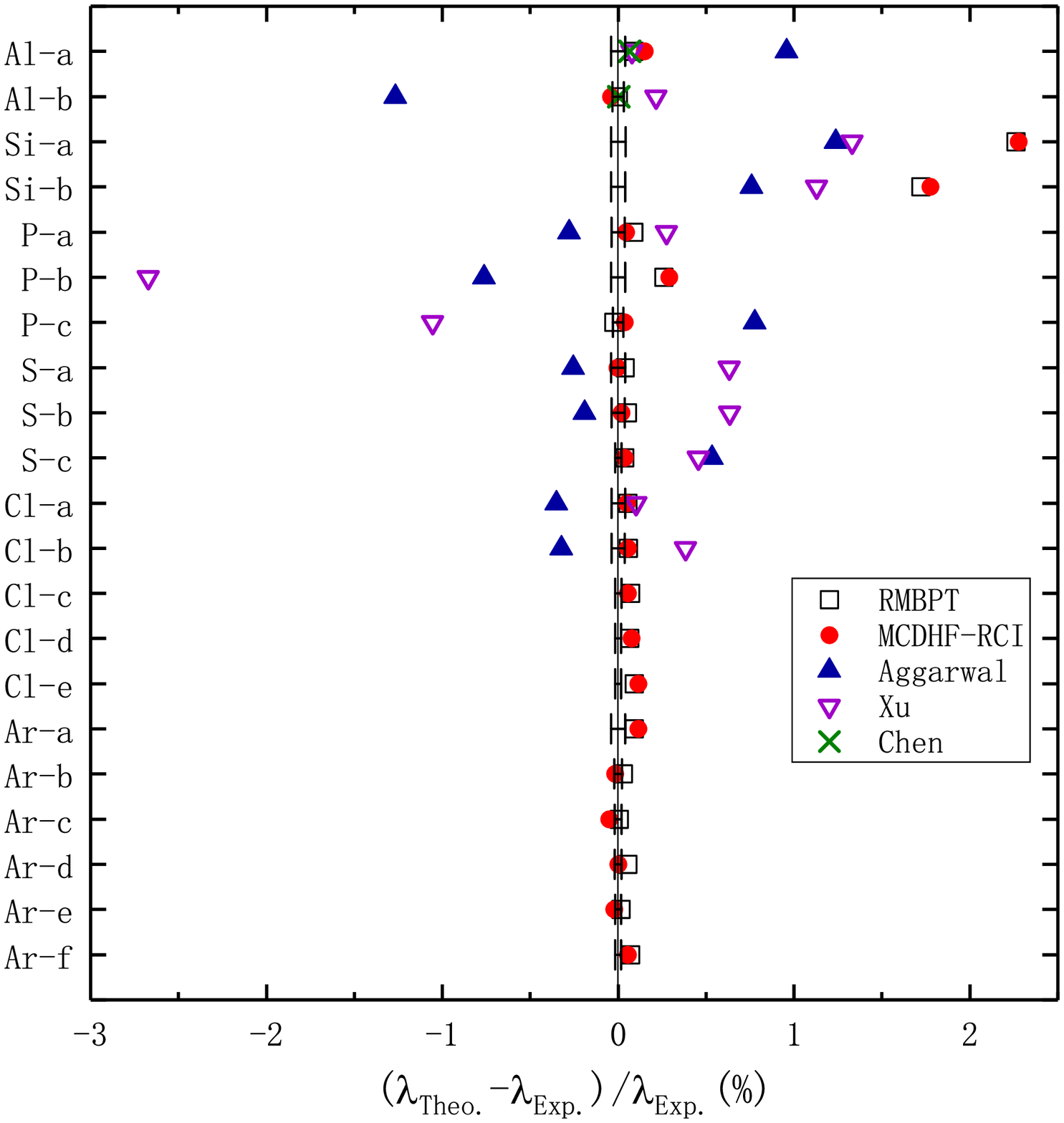}
\caption{\label{fig_JPB2008}The deviations ($\lambda_{\rm{Theo.}}-\lambda_{\rm{Expt.}}$/$\lambda_{\rm{Expt.}}$) in percent  of the different theoretical values  from the experimental wavelengths in Ref.~\cite{Ralchenko.2008.V41.p21003} for the $3p_{3/2} - 3d_{3/2}$, $3s_{1/2} - 3p_{3/2}$, $3s_{1/2} - 3p_{1/2}$ and $3d_{3/2} - 3d_{5/2}$ transitions.
Open square  -- the present RMBPT results; solid circle  -- the present MCDHF-RCI results; solid upper triangle -- Aggarwal and Keenan~\cite{Aggarwal.2014.V100.p1603,Aggarwal.2016.V111-112.p187}; open lower triangle -- Xu $et\ al.$~\cite{Xu.2017.V95.p283}; cross -- Chen and Cheng~\cite{Chen.2011.V84.p12513}. Horizontal error bars are experimental uncertainties. All the corresponding data are available in Table~\ref{Table3pk.tr}.}
\end{figure}

The present RMBPT and MCDHF-RCI wavelengths of W$^{56+}$ $-$ W$^{61+}$ are compared with measured wavelengths in Table~\ref{Table3pk.tr}.
The previous theoretical data are also included in the table, and
we show the deviations $\Delta \lambda$ (in m{\AA}) of the different theoretical values from the experimental wavelengths ~\cite{Clementson.2010.V81.p52509,Lennartsson.2013.V87.p62505,Ralchenko.2008.V41.p21003}. The corresponding deviations in percent ($\lambda_{\rm{Theo.}}-\lambda_{\rm{Expt.}}$/$\lambda_{\rm{Expt.}}$) are presented in Figures~\ref{fig_PRA2010}, \ref{fig_PRA2013}, and \ref{fig_JPB2008}.

It can be seen that the present two data sets (RMBPT and MCDHF-RCI) show an overall better agreement with experimental wavelengths than previous theoretical calculations.	
For example, as shown in Figure~\ref{fig_PRA2010},
our calculated RMBPT and MCDHF-RCI values are evenly scattered on both sides of the experimental values from~\cite{Clementson.2010.V81.p52509}, and deviations between the present results and measured data are within or close to the error bars. The average difference with the standard deviation~\cite{Wang.2017.V229.p37} from measured wavelengths~\cite{Clementson.2010.V81.p52509} are \SI{-0.4}{m{\AA}} $\pm$ \SI{6}{m{\AA}} (\SI{-0.002}{\%} $\pm$ \SI{0.03}{\%}) for the RMBPT results, and \SI{-2}{m{\AA}} $\pm$ \SI{6}{m{\AA}} (\SI{-0.01}{\%} $\pm$ \SI{0.03}{\%}) for the MCDHF-RCI results.
However, most of results from Aggarwal and Keenan~\cite{Aggarwal.2014.V100.p1603,Aggarwal.2016.V111-112.p187} are consistently lower than the measured values~\cite{Clementson.2010.V81.p52509} with an average difference of \SI{-62}{m{\AA}} $\pm$ \SI{13}{m{\AA}} (\SI{-0.3}{\%} $\pm$ \SI{0.06}{\%}).
The Xu $et\ al.$~\cite{Xu.2017.V95.p283} results depart substantially from experimental wavelengths and come with a wide scatter \SI{-32}{m{\AA}} $\pm$ \SI{152}{m{\AA}} (\SI{0.2}{\%} $\pm$ \SI{0.7}{\%}).
The calculations from Chen and Cheng~\cite{Chen.2011.V84.p12513} generally depart from measured lines~\cite{Clementson.2010.V81.p52509} beyond error bars.	

In the experiment~\cite{Lennartsson.2013.V87.p62505}, the FAC-RCI calculations therein were used to aid the line identification. The RCI results were also compared with the measured lines~\citep{Lennartsson.2013.V87.p62505} in Figure~\ref{fig_PRA2013}. As we can see, most of the RCI results from Ref.~\citep{Lennartsson.2013.V87.p62505} differ significantly from the measured values, up to \SI{72}{m{\AA}}, the error bars of the measurement, however, being only from \SI{1}{m{\AA}} to \SI{10}{m{\AA}}. It shows that the RCI calculations in Ref.~\citep{Lennartsson.2013.V87.p62505} are of relatively limited use in the line assignments.
The two lines at $\lambda$ = \SI{35.644(4)}{\AA} and \SI{35.668(4)}{\AA} in the experiment~\cite{Lennartsson.2013.V87.p62505} have not been explicitly identified.
The observed line at $\lambda$ = \SI{35.668(4)}{\AA} might correspond to $3s^23p^5$ $^2P^o_{1/2}$ $\rightarrow$ $3s^23p^5$ $^2P^o_{3/2}$ (an M1 transition of Cl-like W), however,
the calculated RCI wavelength~\cite{Lennartsson.2013.V87.p62505} for this transition is \SI{35.635}{\AA}, which departs largely from the measured wavelength \SI{35.668(4)}{\AA} by
\SI{33}{m{\AA}} beyond the experimental error bar of \SI{4}{m{\AA}}. By comparison, the present MCDHF-RCI and RMBPT values for this transition are, respectively, \SI{35.669}{\AA} and \SI{35.675}{\AA}, that agree well enough with the measured value \SI{35.668(4)}{\AA} to confirm its assignment to the M1 transition $3s^23p^5$ $^2P^o_{1/2}$ $\rightarrow$ $3s^23p^5$ $^2P^o_{3/2}$ of Cl-like W.
Furthermore, the line observed at $\lambda$ = \SI{35.644(4)}{\AA} might correspond to one of the following three transitions: $3s^23p^2$ $^3P_{1}$ $\rightarrow$	$3s^23p^2$ $^3P_{0}$ (an M1 transition of Si-like W), $3s^23p^4(^1S)3d$ $^2D_{5/2}$	$\rightarrow$ $3s^23p^5$ $^2P^o_{3/2}$ (an E1 transition of Cl-like W), and $3s^23p^4$ $^3P_{1}$ $\rightarrow$ $3s^23p^4$ $^3P_{2}$ (an M1 transition of S-like W), with the RCI calculated wavelengths in Ref.~\cite{Lennartsson.2013.V87.p62505} of \SI{35.648}{\AA}, \SI{35.672}{\AA}, and \SI{35.708}{\AA}, respectively. The present RMBPT (MCDHF-RCI) calculations for the M1 transition $3s^23p^2$ $^3P_{1}$ $\rightarrow$	$3s^23p^2$ $^3P_{0}$ are \SI{35.650}{\AA} (\SI{35.657}{\AA}), which agree well with the measured line at $\lambda$ = \SI{35.644(4)}{\AA}, but our RMBPT (MCDHF-RCI) results for the E1 transition $3s^23p^4(^1S)3d$ $^2D_{5/2}$ $\rightarrow$ $3s^23p^5$ $^2P^o_{3/2}$ and the M1 transition of $3s^23p^4$ $^3P_{1}$ $\rightarrow$ $3s^23p^4$ $^3P_{2}$ are \SI{35.673}{\AA} (\SI{35.661}{\AA}) and \SI{35.731}{\AA} (\SI{35.728}{\AA}), respectively, which are markedly different from the measured value. Based on this observation, we suggest to assign the measured line at $\lambda$ = \SI{35.644(4)}{\AA} to the M1 transition $3s^23p^2$ $^3P_{1}$ $\rightarrow$ $3s^23p^2$ $^3P_{0}$ of Si-like W.

Compared with the experimental values from~\cite{Clementson.2010.V81.p52509,Lennartsson.2013.V87.p62505}
(the first two groups of transitions in Table~\ref{Table3pk.tr}), our RMBPT and MCDHF-RCI results are within or close to the experimental uncertainties, and the differences are mostly within \SI{10}{m{\AA}}. However, most of the present results are significantly out of the uncertainties of the experimental values from Ref.~\cite{Ralchenko.2008.V41.p21003} (the last group of transitions in Table~\ref{Table3pk.tr}), and the difference is up to \SI{1.6}{\AA}. We believe that there should be some misidentifications or underestimations of the experimental errors in Ref.~\cite{Ralchenko.2008.V41.p21003}.
From Figure~\ref{fig_JPB2008}, we can see that the two measured lines Si-a ($\lambda=$\SI{71.89(3)}{\AA}, $3s3p^3\  ^2P_2 - 3s^23p^2\ ^3P_1$) and Si-b ($\lambda=$\SI{76.48(3)}{\AA}, $3s3p^3\ ^2P_2 - 3s^23p^2\ ^1P_2$) from~\cite{Ralchenko.2008.V41.p21003} deviate from the present MCDHF-RCI and RMBPT calculations by up to 2\%, but our two data sets agree with each other within 0.01\%. The two lines decay from the same upper level $3s3p^3\ ^2P_2$, but to different lower levels. From our RMBPT and MCDHF-RCI calculations, the splittings between the two lower levels are \SI{74853}{cm}$^{-1}$ and \SI{75307}{cm}$^{-1}$, respectively, but the experimental values from Ref.~\cite{Ralchenko.2008.V41.p21003} gives a splitting of \SI{83483}{cm}$^{-1}$, which gives another evidence that the experimental values need to be re-evaluated.
In addition, the present results are inconsistent with
observation~\cite{Ralchenko.2008.V41.p21003} for other lines, including \SI{76.07(3)}{\AA} of P-like W$^{59+}$, \SI{166.13(3)}{\AA} and \SI{169.11(3)}{\AA} of Cl-like W$^{57+}$, and \SI{173.72(3)}{\AA} of Ar-like W$^{56+}$, by more than an order of magnitude, compared with other measured lines from the same experiment. These lines also deserve further theoretical and experimental studies.
	
\section{Conclusions}\label{conclusions}	
Excitation energies and wavelengths from Al-like W$^{61+}$ to Ar-like W$^{56+}$ ions are calculated and cross-checked by using two state-of-the-art methods, namely, RMBPT and MCDHF-RCI. The two calculations are in excellent agreement with each other on the order of \SI{0.01}{\%}.
The present calculations show a clear improvement of accuracy over previous calculations and achieve spectroscopic accuracy to assist spectroscopists in their assignment and direct identification
of three unidentified lines from~\cite{Lennartsson.2013.V87.p62505}.
The contributions of electron correlation, the Breit interaction, the higher-order frequency-dependent retardation correction, and the leading QED corrections to excitation energies from Al-like W$^{61+}$ to Ar-like W$^{56+}$ ions are studied in detail.
For the corrections beyond the Dirac-Coulomb Hamiltonian, the Breit interaction and QED corrections, respectively, play an important role on the $3p - 3d$ and $3s - 3p$ excitation energies.
The CV correlation involving the $2p$ and $2s$ core electrons and the higher-order retardation corrections beyond the Breit interaction, which were not considered in previous calculations~\cite{Aggarwal.2014.V100.p1603,Aggarwal.2016.V111-112.p187,Xu.2017.V95.p283,Mohan.2014.V92.p177}, should not be ignored for getting high-precision results.
The SE - M2 and SE - M3 methods are more recommended to estimate the SE potential for highly-charged tungsten ions than SE - M1 method.
The complete results incorporating excitation energies, lifetimes, wavelengths, and electric dipole, magnetic dipole, electric quadrupole, and magnetic
quadrupole line strengths, transition rates, and oscillator strengths for the $n = 3$ levels in W$^{56+}$ $-$ W$^{61+}$ will be reported in a future work.
We believe that the present results with spectroscopic accuracy provide precise benchmarks for other theoretical calculations and applications for modeling and diagnosing fusion plasmas.
	
\section*{Acknowledgments}
We acknowledge the support of the National Natural Science Foundation of China (Grant Nos. 12074081, 11974080 and 11703004) and the Nature Science Foundation of Hebei Province, China (A2019201300). KW expresses his gratitude to the support from the visiting researcher program at the Fudan University.
MG and RS acknowledge support from the Belgian FWO \& FNRS Excellence of Science Programme (EOS-O022818F).
PJ acknowledge support from the Swedish
research council under contracts 2015-04842 and 2016-04185.

\onecolumn
\clearpage
\bibliographystyle{model1a-num-names}
\bibliography{reference}

\clearpage
\section*{Figures}

\clearpage

\clearpage

\clearpage

\clearpage
\section*{Tables}
\linespread{1}
\scriptsize
\setlength{\tabcolsep}{3pt}
\begin{longtable}{clllll}
\caption{\label{configuration} Spectral lines studied in this work. For the lines that are measured in~\cite{Clementson.2010.V81.p52509}, we use the original labels consisting of an isoelectronic identifier followed by an integer (Al-1, Al-2, etc.). The lines marked with a Greek letter (Al-$\alpha$, Al-$\beta$, etc.) and an English letter (Al-a, Al-b, etc.) are from~\cite{Lennartsson.2013.V87.p62505} and ~\cite{Ralchenko.2008.V41.p21003}, respectively.
Lower and upper states are presented by their valence configurations, the Ne-like core $1s^2 2s^2 2p^6$ being omitted. Both $LSJ$- and $jj$-couplings are given.}\\
	\toprule
	 & \multicolumn{2}{c}{Lower state}  & \ \ \ \ \ & \multicolumn{2}{c}{Upper state}\\
    \cline{2-3} \cline{5-6}\\
	Line & $LS$-coupling & $jj$-coupling &  & $LS$-coupling & $jj$-coupling \\
    \midrule
	\endfirsthead
	\caption{(\emph{continued})} \\
	\toprule
    & \multicolumn{2}{c}{Lower state}  & \ \ \ \ \ & \multicolumn{2}{c}{Upper state}\\
    \cline{2-3} \cline{5-6}\\
	Line & $LS$-coupling & $jj$-coupling &  & $LS$-coupling & $jj$-coupling \\
    \midrule
	\endhead
    \midrule
	\endfoot
	\bottomrule
	\endlastfoot
Al-1	&$	3s^23p\ ^2P_{1/2}	^o	$&$	(3s^2_{1/2}3p_{1/2})_{1/2}	$& &$	3s^23d\ ^2D_{3/2}		$&$	(3s^2_{1/2}3d_{3/2})_{3/2}	$	\\				
Al-2	&$	3s^23p\ ^2P_{1/2}	^o	$&$	(3s^2_{1/2}3p_{1/2})_{1/2}	$& &$	3s3p^2\ ^2P_{1/2}		$&$	[(3s_{1/2}3p_{1/2})_{1}3p_{3/2}]_{1/2}	$	\\				
Al-3	&$	3s^23p\ ^2P_{1/2}	^o	$&$	(3s^2_{1/2}3p_{1/2})_{1/2}	$& &$	3s3p^2\ ^2D_{3/2}		$&$	[(3s_{1/2}3p_{1/2})_{1}3p_{3/2}]_{3/2}	$	\\				
Al-4	&$	3s^23p\ ^2P_{1/2}	^o	$&$	(3s^2_{1/2}3p_{1/2})_{1/2}	$& &$	3s3p^2\ ^4P_{3/2}		$&$	[(3s_{1/2}3p_{1/2})_{0}3p_{3/2}]_{3/2}	$	\\				
Si-1	&$	3s^23p^2\ ^3P_{0}		$&$	(3s^2_{1/2}3p^2_{1/2})_{0}	$& &$	3s^23p(^2P)3d\ ^3D_{1}	^o	$&$	[(3s^2_{1/2}3p_{1/2})_{1/2}3d_{3/2}]_{1}	$	\\				
Si-2	&$	3s^23p^2\ ^3P_{0}		$&$	(3s^2_{1/2}3p^2_{1/2})_{0}	$& &$	3s3p^3\ ^3D_{1}	^o	$&$	[(3s_{1/2}3p^2_{1/2})_{1/2}3p_{3/2}]_{1}	$	\\				
P-1	&$	3s^23p^3\ ^2P_{3/2}  	^o	$&$	[(3s^2_{1/2}3p^2_{1/2})_03p_{3/2}]_{3/2}	$& &$	3s^23p^2(^1D)3d\ ^2D_{3/2} 		$&$	\{[(3s^2_{1/2}3p_{1/2})_{1/2}3p_{3/2}]_{2}3d_{3/2}\}_{3/2}	$	\\				
P-2	&$	3s^23p^3\ ^2P_{3/2}  	^o	$&$	[(3s^2_{1/2}3p^2_{1/2})_03p_{3/2}]_{3/2}	$& &$	3s^23p^2(^3P)3d\ ^2D_{5/2} 		$&$	\{[(3s^2_{1/2}3p_{1/2})_{1/2}3p_{3/2}]_{1}3d_{3/2}\}_{5/2}	$	\\				
P-3	&$	3s^23p^3\ ^2P_{3/2}  	^o	$&$	[(3s^2_{1/2}3p^2_{1/2})_03p_{3/2}]_{3/2}	$& &$	3s3p^4\ ^4P_{5/2} 		$&$	[(3s_{1/2}3p^2_{1/2})_{1/2}3p^2_{3/2}]_{5/2}	$	\\				
S-1	&$	     3s^23p^4\ ^3P_{2}      		$&$	[(3s^2_{1/2}3p^2_{1/2})_03p^2_{3/2}]_{2}	$& &$	3s^2 3p^3(^2D)3d\ ^3S_{1}  	^o	$&$	\{[(3s^2_{1/2}3p_{1/2})_{1/2}3p^2_{3/2}]_{5/2}3d_{3/2}\}_{1}	$	\\				
S-2	&$	     3s^23p^4\ ^3P_{2}      		$&$	[(3s^2_{1/2}3p^2_{1/2})_03p^2_{3/2}]_{2}	$& &$	3s^2 3p^3(^2D)3d\ ^3P_{2}    	^o	$&$	\{[(3s^2_{1/2}3p_{1/2})_{1/2}3p^2_{3/2}]_{5/2}3d_{3/2}\}_{2}	$	\\				
S-3	&$	     3s^23p^4\ ^3P_{2}      		$&$	[(3s^2_{1/2}3p^2_{1/2})_03p^2_{3/2}]_{2}	$& &$	3s^2 3p^3(^2D)3d\ ^1F_{3}  	^o	$&$	\{[(3s^2_{1/2}3p_{1/2})_{1/2}3p^2_{3/2}]_{5/2}3d_{3/2}\}_{3}	$	\\				
S-4	&$	     3s^23p^4\ ^3P_{2}      		$&$	[(3s^2_{1/2}3p^2_{1/2})_03p^2_{3/2}]_{2}	$& &$	   3s3p^5\ ^3P_{2}    	^o	$&$	[(3s_{1/2}3p^2_{1/2})_{1/2}3p^3_{3/2}]_{2}	$	\\				
Cl-1	&$	3s^23p^5\ ^2P_{3/2}	^o	$&$	[(3s^2_{1/2}3p^2_{1/2})_03p^3_{3/2}]_{3/2}	$& &$	3s^23p^4(^1D)3d\ ^2S_{1/2}		$&$	\{[(3s^2_{1/2}3p_{1/2})_{1/2}3p^3_{3/2}]_{2}3d_{3/2}\}_{1/2}	$	\\				
Cl-2	&$	3s^23p^5\ ^2P_{3/2}	^o	$&$	[(3s^2_{1/2}3p^2_{1/2})_03p^3_{3/2}]_{3/2}	$& &$	3s^23p^4(^1D)3d\ ^2P_{3/2}		$&$	\{[(3s^2_{1/2}3p_{1/2})_{1/2}3p^3_{3/2}]_{2}3d_{3/2}\}_{3/2}	$	\\				
Cl-3	&$	3s^23p^5\ ^2P_{3/2}	^o	$&$	[(3s^2_{1/2}3p^2_{1/2})_03p^3_{3/2}]_{3/2}	$& &$	3s^23p^4(^1D)3d\ ^2F_{5/2}		$&$	\{[(3s^2_{1/2}3p_{1/2})_{1/2}3p^3_{3/2}]_{2}3d_{3/2}\}_{5/2}	$	\\				
Ar-1	&$	3s^23p^6\ ^1S_{0} 		$&$	[(3s^2_{1/2}3p^2_{1/2})_03p^4_{3/2}]_{0}	$& &$	3s^23p^5(^2P)3d\ ^3D_{1} 	^o	$&$	\{[(3s^2_{1/2}3p_{1/2})_{1/2}3p^4_{3/2}]_{1/2}\}3d_{3/2}\}_{1}	$	\\				
	&$			$&$		$& &$			$&$		$	\\				
Al-$\alpha$	&$	3s^23p\ ^2P_{1/2}	^o	$&$	(3s^2_{1/2}3p_{1/2})_{1/2}	$& &$	3s^23p\ ^2P_{3/2}	^o	$&$	(3s^2_{1/2}3p_{3/2})_{3/2}	$	\\				
Al-$\beta$	&$	3s^23p\ ^2P_{3/2}	^o	$&$	(3s^2_{1/2}3p_{3/2})_{3/2}	$& &$	3s^23d\ ^2D_{5/2}		$&$	(3s^2_{1/2}3d_{5/2})_{5/2}	$	\\				
Si-$\alpha$	&$	3s^23p^2\ ^3P_{0}		$&$	(3s^2_{1/2}3p^2_{1/2})_{0}	$& &$	3s^23p^2\ ^1D_{2}		$&$	[(3s^2_{1/2}3p_{1/2})_{1/2}3p_{3/2}]_{2}	$	\\				
Si-$\beta$	&$	3s3p^3\ ^3P_{2}	^o	$&$	\{[3s_{1/2}3p^2_{1/2}]_{1/2}3p_{3/2}\}_{2}	$& &$	3s3p^2(^4P)3d\ ^3F_{2}		$&$	[(3s_{1/2}3p^2_{1/2})_{1/2}3d_{5/2}]_{2}	$	\\				
Si-$\gamma$	&$	3s^23p^2\ ^1D_{2}		$&$	[(3s^2_{1/2}3p_{1/2})_{0}3p_{3/2}]_{2}	$& &$	3s^23p(^2P)3d\ ^3F_{3}	^o	$&$	[(3s^2_{1/2}3p_{1/2})_{1/2}3d_{5/2}]_{3}	$	\\				
Si-$\delta$	&$	3s3p^3\ ^3D_{1}	^o	$&$	\{[3s_{1/2}3p^2_{1/2}]_{1/2}3p_{3/2}\}_{1}	$& &$	3s3p^2(^4P)3d\ ^3F_{2}		$&$	[(3s_{1/2}3p^2_{1/2})_{1/2}3d_{5/2}]_{2}	$	\\				
P-$\alpha$	&$	3s^23p^3\ ^2P_{3/2}  	^o	$&$	[(3s^2_{1/2}3p^2_{1/2})_03p_{3/2}]_{3/2}	$& &$	3s^23p^3\ ^2D_{5/2} 	^o	$&$	[(3s^2_{1/2}3p_{1/2})_{1/2}3p^2_{3/2}]_{5/2}	$	\\				
P-$\beta$	&$	3s^23p^3\ ^2P_{3/2}  	^o	$&$	[(3s^2_{1/2}3p^2_{1/2})_03p_{3/2}]_{3/2}	$& &$	3s^23p^3\ ^4S_{3/2} 	^o	$&$	[(3s^2_{1/2}3p_{1/2})_{1/2}3p^2_{3/2}]_{3/2}	$	\\				
P-$\gamma$	&$	3s^23p^3\ ^2P_{3/2}  	^o	$&$	[(3s^2_{1/2}3p^2_{1/2})_03p_{3/2}]_{3/2}	$& &$	3s^23p^2(^1S)3d\ ^2D_{5/2} 		$&$	[(3s^2_{1/2}3p^2_{1/2})_03d_{5/2}]_{5/2}	$	\\				
S-$\alpha$	&$	     3s^23p^4\ ^3P_{2}      		$&$	[(3s^2_{1/2}3p^2_{1/2})_03p^2_{3/2}]_{2}	$& &$	   3s^23p^3(^2P)3d\ ^3D_{3}  	^o	$&$	\{[(3s^2_{1/2}3p^2_{1/2})_03p_{3/2}]_{3/2}3d_{5/2}\}_{3}	$	\\				
S-$\beta$	&$	 3s^23p^4\ ^1S_{0}      		$&$	[(3s^2_{1/2}3p^2_{1/2})_03p^2_{3/2}]_{0}	$& &$	 3s^23p^3(^2P)3d\ ^1P_{1}  	^o	$&$	\{[(3s^2_{1/2}3p^2_{1/2})_03p_{3/2}]_{3/2}3d_{5/2}\}_{1}	$	\\				
S-$\gamma$	&$	     3s^23p^4\ ^3P_{2}      		$&$	[(3s^2_{1/2}3p^2_{1/2})_03p^2_{3/2}]_{2}	$& &$	    3s^23p^3(^2P)3d\ ^3P_{2}    	^o	$&$	\{[(3s^2_{1/2}3p^2_{1/2})_03p_{3/2}]_{3/2}3d_{5/2}\}_{2}	$	\\				
Cl-$\alpha$	&$	3s^23p^5\ ^2P_{3/2}	^o	$&$	[(3s^2_{1/2}3p^2_{1/2})_03p^3_{3/2}]_{3/2}	$& &$	3s^23p^4(^3P)3d\ ^2D_{5/2}		$&$	\{[(3s^2_{1/2}3p^2_{1/2})_03p^2_{3/2}]_{2}3d_{5/2}\}_{5/2}	$	\\				
Cl-$\beta$	&$	3s^23p^5\ ^2P_{3/2}	^o	$&$	[(3s^2_{1/2}3p^2_{1/2})_03p^3_{3/2}]_{3/2}	$& &$	3s^23p^4(^3P)3d\ ^4P_{3/2}		$&$	\{[(3s^2_{1/2}3p^2_{1/2})_03p^2_{3/2}]_{2}3d_{5/2}\}_{3/2}	$	\\				
Ar-$\alpha$	&$	3s^23p^5(^2P)3d\ ^3F_{3} 	^o	$&$	\{[(3s^2_{1/2}3p^2_{1/2})_03p^3_{3/2}]_{3/2}3d_{3/2}\}_{3}	$& &$	3s^23p^4(^3P)3d^2\ ^3G_{4} 		$&$	\{[(3s^2_{1/2}3p^2_{1/2})_{0}3p^2_{3/2}]_{0}\}3d_{3/2}3d_{5/2}\}_{4}	$	\\				
Ar-$\beta$	&$	3s^23p^6\ ^1S_{0} 		$&$	[(3s^2_{1/2}3p^2_{1/2})_03p^4_{3/2}]_{0}	$& &$	3s^23p^5(^2P)3d\ ^1P_{1} 	^o	$&$	\{[(3s^2_{1/2}3p^2_{1/2})_03p^3_{3/2}]_{3/2}3d_{5/2}\}_{1}	$	\\				
Ar-$\gamma$	&$	3s^23p^5(^2P)3d\ ^3F_{3} 	^o	$&$	\{[(3s^2_{1/2}3p^2_{1/2})_03p^3_{3/2}]_{3/2}3d_{3/2}\}_{3}	$& &$	3s^23p^4(^1D)3d^2\ ^3G_{3} 		$&$	\{[(3s^2_{1/2}3p^2_{1/2})_03p^2_{3/2}]_{2}3d_{3/2}3d_{5/2}\}_{3}	$	\\				
	&$			$&$		$& &$			$&$		$	\\				
Al-a	&$	3s3p^2\ ^4P_{1/2}   		$&$	(3s_{1/2}3p^2_{1/2})_{1/2}	$& &$	3s^23p\ ^2P_{3/2}	^o	$&$	(3s^2_{1/2}3p_{3/2})_{3/2}	$	\\				
Al-b	&$	3s^23p\ ^2P_{1/2}	^o	$&$	(3s^2_{1/2}3p_{1/2})_{1/2}	$& &$	3s3p^2\ ^4P_{1/2}   		$&$	(3s_{1/2}3p^2_{1/2})_{1/2}	$	\\				
Si-a	&$	3s^23p^2\ ^3P_{1}		$&$	[(3s^2_{1/2}3p_{1/2})_{1/2}3p_{3/2}]_{1}	$& &$	3s3p^3\ ^3P_{2}	^o	$&$	\{[3s_{1/2}3p^2_{1/2}]_{1/2}3p_{3/2}\}_{2}	$	\\				
Si-b	&$	3s^23p^2\ ^1D_{2}		$&$	[(3s^2_{1/2}3p_{1/2})_{1/2}3p_{3/2}]_{2}	$& &$	3s3p^3\ ^3P_{2}	^o	$&$	\{[3s_{1/2}3p^2_{1/2}]_{1/2}3p_{3/2}\}_{2}	$	\\				
P-a	&$	3s^23p^3\ ^2P_{3/2}  	^o	$&$	[(3s^2_{1/2}3p^2_{1/2})_03p_{3/2}]_{3/2}	$& &$	3s^23p^2(^3P)3d\ ^4F_{3/2} 		$&$	[(3s^2_{1/2}3p^2_{1/2})_03d_{3/2}]_{3/2}	$	\\				
P-b	&$	3s^23p^3\ ^2D_{5/2} 	^o	$&$	[(3s^2_{1/2}3p_{1/2})_{1/2}3p^2_{3/2}]_{5/2}	$& &$	3s3p^4\ ^4P_{5/2} 		$&$	[(3s_{1/2}3p^2_{1/2})_{1/2}3p^2_{3/2}]_{5/2}	$	\\				
P-c	&$	3s^23p^2(^3P)3d\ ^4F_{3/2} 		$&$	[(3s^2_{1/2}3p^2_{1/2})_03d_{3/2}]_{3/2}	$& &$	3s^23p^3\ ^2D_{5/2} 	^o	$&$	[(3s^2_{1/2}3p_{1/2})_{1/2}3p^2_{3/2}]_{5/2}	$	\\				
S-a	&$	     3s^23p^4\ ^3P_{2}      		$&$	[(3s^2_{1/2}3p^2_{1/2})_03p^2_{3/2}]_{2}	$& &$	  3s^23p^3(^2P)3d\ ^3F_{3}  	^o	$&$	\{[(3s^2_{1/2}3p^2_{1/2})_03p_{3/2}]_{3/2}3d_{3/2}\}_{3}	$	\\				
S-b	&$	     3s^23p^4\ ^3P_{2}      		$&$	[(3s^2_{1/2}3p^2_{1/2})_03p^2_{3/2}]_{2}	$& &$	   3s^23p^3(^2P)3d\ ^3D_{2}    	^o	$&$	\{[(3s^2_{1/2}3p^2_{1/2})_03p_{3/2}]_{3/2}3d_{3/2}\}_{2}	$	\\				
S-c	&$	  3s^23p^3(^2P)3d\ ^3F_{3}  	^o	$&$	\{[(3s^2_{1/2}3p^2_{1/2})_03p_{3/2}]_{3/2}3d_{3/2}\}_{3}	$& &$	    3s^23p^3(^2P)3d\ ^3F_{4}    	^o	$&$	\{[(3s^2_{1/2}3p^2_{1/2})_03p_{3/2}]_{3/2}3d_{5/2}\}_{4}	$	\\				
Cl-a	&$	3s^23p^5\ ^2P_{3/2}	^o	$&$	[(3s^2_{1/2}3p^2_{1/2})_03p^3_{3/2}]_{3/2}	$& &$	3s^23p^4(^3P)3d\ ^4D_{5/2}		$&$	\{[(3s^2_{1/2}3p^2_{1/2})_03p^2_{3/2}]_{2}3d_{3/2}\}_{5/2}	$	\\				
Cl-b	&$	3s^23p^5\ ^2P_{3/2}	^o	$&$	[(3s^2_{1/2}3p^2_{1/2})_03p^3_{3/2}]_{3/2}	$& &$	3s^23p^4(^3P)3d\ ^4D_{3/2}		$&$	\{[(3s^2_{1/2}3p^2_{1/2})_03p^2_{3/2}]_{2}3d_{3/2}\}_{3/2}	$	\\				
Cl-c	&$	3s^23p^4(^3P)3d\ ^4D_{5/2}		$&$	\{[(3s^2_{1/2}3p^2_{1/2})_03p^2_{3/2}]_{2}3d_{3/2}\}_{5/2}	$& &$	3s^23p^4(^3P)3d\ ^4D_{7/2}		$&$	\{[(3s^2_{1/2}3p^2_{1/2})_03p^2_{3/2}]_{2}3d_{5/2}\}_{7/2}	$	\\				
Cl-d	&$	3s^23p^4(^3P)3d\ ^2F_{7/2}		$&$	\{[(3s^2_{1/2}3p^2_{1/2})_03p^2_{3/2}]_{2}3d_{3/2}\}_{7/2}	$& &$	3s^23p^4(^3P)3d\ ^4F_{9/2}		$&$	\{[(3s^2_{1/2}3p^2_{1/2})_03p^2_{3/2}]_{2}3d_{5/2}\}_{9/2}	$	\\				
Ar-a	&$	3s^23p^6\ ^1S_{0} 		$&$	[(3s^2_{1/2}3p^2_{1/2})_03p^4_{3/2}]_{0}	$& &$	3s^23p^5(^2P)3d\ ^3P_{1} 	^o	$&$	\{[(3s^2_{1/2}3p^2_{1/2})_03p^3_{3/2}]_{3/2}3d_{3/2}\}_{1}	$	\\				
Ar-b	&$	3s^23p^5(^2P)3d\ ^3F_{3} 	^o	$&$	\{[(3s^2_{1/2}3p^2_{1/2})_03p^3_{3/2}]_{3/2}3d_{3/2}\}_{3}	$& &$	3s^23p^5(^2P)3d\ ^3D_{3} 	^o	$&$	\{[(3s^2_{1/2}3p^2_{1/2})_03p^3_{3/2}]_{3/2}3d_{5/2}\}_{3}	$	\\				
Ar-c	&$	3s^23p^5(^2P)3d\ ^3D_{2} 	^o	$&$	\{[(3s^2_{1/2}3p^2_{1/2})_03p^3_{3/2}]_{3/2}3d_{3/2}\}_{2}	$& &$	3s^23p^5(^2P)3d\ ^3D_{3} 	^o	$&$	\{[(3s^2_{1/2}3p^2_{1/2})_03p^3_{3/2}]_{3/2}3d_{5/2}\}_{3}	$	\\				
Ar-d	&$	3s^23p^5(^2P)3d\ ^3P_{1} 	^o	$&$	\{[(3s^2_{1/2}3p^2_{1/2})_03p^3_{3/2}]_{3/2}3d_{3/2}\}_{1}	$& &$	3s^23p^5(^2P)3d\ ^1D_{2} 	^o	$&$	\{[(3s^2_{1/2}3p^2_{1/2})_03p^3_{3/2}]_{3/2}3d_{5/2}\}_{2}	$	\\				
Ar-e	&$	3s^23p^5(^2P)3d\ ^3D_{2} 	^o	$&$	\{[(3s^2_{1/2}3p^2_{1/2})_03p^3_{3/2}]_{3/2}3d_{3/2}\}_{2}	$& &$	3s^23p^5(^2P)3d\ ^1D_{2} 	^o	$&$	\{[(3s^2_{1/2}3p^2_{1/2})_03p^3_{3/2}]_{3/2}3d_{5/2}\}_{2}	$	\\				
Ar-f	&$	3s^23p^5(^2P)3d\ ^3F_{3} 	^o	$&$	\{[(3s^2_{1/2}3p^2_{1/2})_03p^3_{3/2}]_{3/2}3d_{3/2}\}_{3}	$& &$	3s^23p^5(^2P)3d\ ^3F_{4} 	^o	$&$	\{[(3s^2_{1/2}3p^2_{1/2})_03p^3_{3/2}]_{3/2}3d_{5/2}\}_{4}	$	\\				
\end{longtable}

\clearpage
\linespread{1}
\tiny
\setlength{\tabcolsep}{3pt}
\begin{longtable}{clccccccccccrrrrrrrr}
\caption{\label{Table3pk_E-NIST} Excitation energies ($E$, in cm$^{-1}$) from the present MCDHF-RCI calculations compared with the experimental values (Expt)~\cite{Clementson.2010.V81.p52509}. The estimated uncertainty of the experimental value for each level is reported in brackets in the Expt column. The MCDHF-RCI values are calculated using the Dirac-Coulomb Hamitonian with CSF expansions targeting valence (VV) and core-valence (CV) electron correlation; The DCB and DCTP values are obtained by considering in the RCI step the Dirac-Coulomb-Breit and Dirac-Coulomb-Transverse-Photon Hamiltonians, respectively; the column labeled +QED list the final MCDHF-RCI results which further include the QED corrections that are estimated using the three different models (M1-M3). The present RMBPT values are also included. The corresponding differences $\Delta E_{\rm{Theo.}}$=$E_{\rm{Theo.}}$-$E_{\rm{Expt.}}$ are reported (in cm$^{-1}$).}\\
\toprule				
\multirow{4}{*}{Ion} & \multirow{4}{*}{Level}  &  \multicolumn{8}{c}{$E$ (cm$^{-1}$)} & & \multicolumn{8}{c}{$\Delta E$ (cm$^{-1}$)}  \\
\cline{3-11} \cline{13-20}\\
& & Expt. &  \multicolumn{7}{c}{MCDHF-RCI}  & RMBPT & & \multicolumn{7}{c}{MCDHF-RCI} & RMBPT \\
\cline{4-10} \cline{13-19}\\
& &  &   &  &  &  &  \multicolumn{3}{c}{+QED}  &  & &  &  &  &  &\multicolumn{3}{c}{+QED}& \\
\cline{8-10} \cline{17-19}\\
& &  &  VV & +CV & DCB & DCTP & M1  & M2 & M3 &  & & VV & +CV & DCB & DCTP & M1&M2&M3& \\
\midrule
\endhead
\endfoot
\bottomrule
\endlastfoot
					
Al-1	&		$	3s^23d\ ^2D_{3/2}	$	&	4817900	(900)	&			4870222 	&		4866567 	&		4830456 	&		4829017 							&		4817771 	&	4817434 	&	4817188 			&	4817420 				&	&	52322 	&	48667 	&	12556 	&	11117 	&					-129 	&	-466 	&	-712 		&	-480 				\\
Al-2	&		$	3s3p^2\ ^2P_{1/2}	$	&	4436000	(800)	&			4476119 	&		4473687 	&		4476365 	&		4471851 							&		4433928 	&	4435298 	&	4435377 			&	4435377 				&	&	40119 	&	37687 	&	40365 	&	35851 	&					-2072 	&	-702 	&	-623 		&	-623 				\\
Al-3	&		$	3s3p^2\ ^2D_{3/2}	$	&	4355200	(1100)	&			4396268 	&		4396591 	&		4388449 	&		4384731 							&		4353197 	&	4354193 	&	4354263 			&	4354452 				&	&	41068 	&	41391 	&	33249 	&	29531 	&					-2003 	&	-1007 	&	-937 		&	-748 				\\
Al-4	&		$	3s3p^2\ ^4P_{3/2}	$	&	4035500	(1600)	&			4072599 	&		4075437 	&		4077816 	&		4073311 							&		4036250 	&	4037609 	&	4037631 			&	4038120 				&	&	37099 	&	39937 	&	42316 	&	37811 	&					750 	&	2109 	&	2131 		&	2620 				\\
Si-1	&		$	3s^23p(^2P)3d\ ^3D^o_{1}	$	&	4933400	(1200)	&			4990401 	&		4981000 	&		4944455 	&		4943140 							&		4934251 	&	4933889 	&	4933886 			&	4932912 				&	&	57001 	&	47600 	&	11055 	&	9740 	&					851 	&	489 	&	486 		&	-488 				\\
Si-2	&		$	3s3p^3\ ^3D^o_{1}	$	&	4387300	(1000)	&			4430110 	&		4427816 	&		4420759 	&		4417018 							&		4387010 	&	4388156 	&	4388275 			&	4388082 				&	&	42810 	&	40516 	&	33459 	&	29718 	&					-290 	&	856 	&	975 		&	782 				\\
P-1	&		$	3s^23p^2(^1D)3d\ ^2D_{3/2} 	$	&	5010300	(2000)  	&			5062666 	&		5055915 	&		5022286 	&		5020789 							&		5010353 	&	5010133 	&	5010020 			&	5009016 				&	&	52366 	&	45615 	&	11986 	&	10489 	&					53 	&	-167 	&	-280 		&	-1284 				\\
P-2	&		$	3s^23p^2(^3P)3d\ ^2D_{5/2} 	$	&	4921500	(1200)  	&			4975546 	&		4971315 	&		4927867 	&		4927209 							&		4923182 	&	4922537 	&	4922471 			&	4921502 				&	&	54046 	&	49815 	&	6367 	&	5709 	&					1682 	&	1037 	&	971 		&	2 				\\
P-3	&		$	3s3p^4\ ^4P_{5/2} 	$	&	4159400	(1000) 	&			4199473 	&		4199361 	&		4194307 	&		4190186 							&		4157124 	&	4158499 	&	4158523 			&	4158869 				&	&	40073 	&	39961 	&	34907 	&	30786 	&					-2276 	&	-901 	&	-877 		&	-531 				\\
S-1	&		$	3s^2 3p^3(^2D)3d\ ^3S^o_{1}  	$	&	5062800	(2100)	&			5114885 	&		5109250 	&		5074042 	&		5073037 							&		5066102 	&	5065599 	&	5065600 			&	5064317 				&	&	52085 	&	46450 	&	11242 	&	10237 	&					3302 	&	2799 	&	2800 		&	1517 				\\
S-2	&		$	3s^2 3p^3(^2D)3d\ ^3P^o_{2}    	$	&	5018300	(2500)	&			5071977 	&		5066689 	&		5028431 	&		5027550 							&		5021539 	&	5020965 	&	5021089 			&	5019828 				&	&	53677 	&	48389 	&	10131 	&	9250 	&					3239 	&	2665 	&	2789 		&	1528 				\\
S-3	&		$	3s^2 3p^3(^2D)3d\ ^1F^o_{3}  	$	&	4963500	(1500)	&			5017897 	&		5013815 	&		4968599 	&		4968098 							&		4964990 	&	4964223 	&	4964257 			&	4963025 				&	&	54397 	&	50315 	&	5099 	&	4598 	&					1490 	&	723 	&	757 		&	-475 				\\
S-4	&		$	   3s3p^5\ ^3P^o_{2}    	$	&	4282700	(1800)	&			4322043 	&		4320956 	&		4315574 	&		4311808 							&		4281212 	&	4282365 	&	4282472 			&	4282655 				&	&	39343 	&	38256 	&	32874 	&	29108 	&					-1488 	&	-335 	&	-228 		&	-45 				\\
Cl-1	&		$	3s^23p^4(^1D)3d\ ^2S_{1/2}	$	&	5097000 	(3000) 	&			5150230 	&		5142605 	&		5109247 	&		5107989 							&		5098691 	&	5098485 	&	5098399 			&	5097879 				&	&	53230 	&	45605 	&	12247 	&	10989 	&					1691 	&	1485 	&	1399 		&	879 				\\
Cl-2	&		$	3s^23p^4(^1D)3d\ ^2P_{3/2}	$	&	5046900 	(2300) 	&			5095809 	&		5090958 	&		5049389 	&		5048891 							&		5045735 	&	5045115 	&	5045154 			&	5044136 				&	&	48909 	&	44058 	&	2489 	&	1991 	&					-1165 	&	-1785 	&	-1746 		&	-2764 				\\
Cl-3	&		$	3s^23p^4(^1D)3d\ ^2F_{5/2}	$	&	5024100 	(2500) 	&			5078662 	&		5073745 	&		5029442 	&		5028963 							&		5025765 	&	5025147 	&	5025126 			&	5024368 				&	&	54562 	&	49645 	&	5342 	&	4863 	&					1665 	&	1047 	&	1026 		&	268 				\\
Ar-1	&		$	3s^23p^5(^2P)3d\ ^3D^o_{1} 	$	&	5081600	(1800)	&			5134064 	&		5128844 	&		5086476 	&		5086025 							&		5083128 	&	5082465 	&	5082334 			&	5082076 				&	&	52464 	&	47244 	&	4876 	&	4425 	&					1528 	&	865 	&	734 		&	476 				\\

\end{longtable}

\clearpage
\linespread{1}
\tiny
\setlength{\tabcolsep}{3pt}
\begin{longtable}{ccccccccccccrrrrrrrr}
\caption{\label{Table3pk.tr} Computed wavelengths (in {\AA}) from the present RMBPT and MCDHF-RCI using the third QED potential (M3), as well as from the previous works (Aggarwal: from Aggarwal and Keenan~\cite{Aggarwal.2014.V100.p1603,Aggarwal.2016.V111-112.p187}; Xu: from Xu $et\ al.$~\cite{Xu.2017.V95.p283}; Chen: from  Chen and Cheng~\cite{Chen.2011.V84.p12513}; Ekman: from Ekman $et\ al.$~\cite{Ekman.2018.V120.p152}; Safronova: Safronova $et\ al.$~\cite{Safronova.2010.V43.p74026}; Mohan: Mohan $et\ al.$~\cite{Mohan.2014.V92.p177}),   with the measured values from~\cite{Ralchenko.2008.V41.p21003,Lennartsson.2013.V87.p62505,Clementson.2010.V81.p52509}.
The deviations $\Delta \lambda$ (in m{\AA}) of the different theoretical values from the experimental wavelengths are also listed. The number reported in parenthesis, after the experimental wavelength, is the estimated experimental uncertainty.}\\
\toprule           			
&  \multicolumn{10}{c}{$\lambda$({\AA})} & & \multicolumn{8}{c}{$\Delta \lambda$(m{\AA})}  \\
\cline{3-11} \cline{13-20}\\
Line &  TM & Expt. & RMBPT & MCDHF-RCI & Aggarwal & Xu &Chen & Ekman & Safronova & Mohan &  & RMBPT & MCDHF-RCI & Aggarwal & Xu &Chen & Ekman & Safronova & Mohan\\
\midrule
	\endhead
	\endfoot
	\bottomrule
	\endlastfoot
Al-1	&	E1	&	20.756	(4)	&	20.758 	&	20.759 	&	20.71 	&	20.744 	&	20.78 	&	20.750 	&	20.77 	&		&		&	2 	&	3 	&	-46 	&	-12 	&	24 	&	-6 	&	14 	&		\\							
Al-2	&	E1	&	22.543	(4)	&	22.546 	&	22.546 	&	22.46 	&	22.556 	&	22.54 	&	22.527 	&	22.55 	&		&		&	3 	&	3 	&	-83 	&	13 	&	-3 	&	-16 	&	7 	&		\\							
Al-3	&	E1	&	22.961	(6)	&	22.965 	&	22.966 	&	22.90 	&	22.991 	&	22.97 	&	22.948 	&	22.98 	&		&		&	4 	&	5 	&	-61 	&	30 	&	9 	&	-13 	&	19 	&		\\							
Al-4	&	E1	&	24.780	(10)	&	24.764 	&	24.767 	&	24.71 	&	24.790 	&	24.77 	&	24.743 	&	24.78 	&		&		&	-16 	&	-13 	&	-70 	&	10 	&	-10 	&	-37 	&	0 	&		\\							
Si-1	&	E1	&	20.270	(5)	&	20.272 	&	20.268 	&	20.21 	&	20.629 	&	20.24 	&		&		&		&		&	2 	&	-2 	&	-60 	&	359 	&	-30 	&		&		&		\\							
Si-2	&	E1	&	22.793	(5)	&	22.789 	&	22.788 	&	22.72 	&	22.690 	&	22.78 	&		&		&		&		&	-4 	&	-5 	&	-73 	&	-103 	&	-13 	&		&		&		\\							
P-1	&	E1	&	19.959	(8)	&	19.964 	&	19.960 	&	19.89 	&	20.214 	&		&		&		&		&		&	5 	&	1 	&	-69 	&	255 	&		&		&		&		\\							
P-2	&	E1	&	20.319	(5)	&	20.319 	&	20.315 	&	20.27 	&	20.402 	&		&		&		&		&		&	0 	&	-4 	&	-49 	&	83 	&		&		&		&		\\							
P-3	&	E1	&	24.042	(9)	&	24.045 	&	24.047 	&	23.97 	&	23.780 	&		&		&		&		&		&	3 	&	5 	&	-72 	&	-262 	&		&		&		&		\\							
S-1	&	E1	&	19.752	(8)	&	19.746 	&	19.741 	&	19.69 	&		&		&		&		&		&		&	-6 	&	-11 	&	-62 	&		&		&		&		&		\\							
S-2	&	E1	&	19.927	(10)	&	19.921 	&	19.916 	&	19.87 	&	20.146 	&		&		&		&		&		&	-6 	&	-11 	&	-57 	&	219 	&		&		&		&		\\							
S-3	&	E1	&	20.147	(6)	&	20.149 	&	20.144 	&	20.11 	&	20.106 	&		&		&		&		&		&	2 	&	-3 	&	-37 	&	-41 	&		&		&		&		\\							
S-4	&	E1	&	23.350	(10)	&	23.350 	&	23.351 	&	23.28 	&	23.383 	&		&		&		&		&		&	0 	&	1 	&	-70 	&	33 	&		&		&		&		\\							
Cl-1	&	E1	&	19.62	(1)	&	19.616 	&	19.614 	&	19.54 	&	19.607 	&		&		&		&	19.5 	&		&	-4 	&	-6 	&	-80 	&	-13 	&		&		&		&	-120 	\\							
Cl-2	&	E1	&	19.814	(9)	&	19.825 	&	19.821 	&	19.77 	&	19.781 	&		&		&		&	19.7 	&		&	11 	&	7 	&	-44 	&	-33 	&		&		&		&	-114 	\\							
Cl-3	&	E1	&	19.904	(10)	&	19.903 	&	19.900 	&	19.85 	&	19.844 	&		&		&		&	19.8 	&		&	-1 	&	-4 	&	-54 	&	-60 	&		&		&		&	-104 	\\							
Ar-1	&	E1	&	19.679	(7)	&	19.677 	&	19.676 	&		&		&	19.65 	&		&		&		&		&	-2 	&	-3 	&		&		&	-29 	&		&		&		\\							
	&					&		&		&		&		&		&		&		&		&		&		&		&		&		&		&		&		&		\\							
Al-$\alpha$	&	M1	&	34.110	(7)	&	34.111 	&	34.117 	&	34.12 	&	34.136 	&	34.10 	&	34.058 	&	34.12 	&		&		&	1 	&	7 	&	10 	&	26 	&	-10 	&	-52 	&	10 	&		\\							
Al-$\beta$	&	E1	&	40.37	(1)	&	40.367 	&	40.364 	&	40.29 	&		&		&	40.419 	&	40.40 	&		&		&	-3 	&	-6 	&	-80 	&		&		&	49 	&	30 	&		\\							
Si-$\alpha$	&	E2	&	34.720	(1)	&	34.723 	&	34.725 	&	34.74 	&	34.665 	&		&		&		&		&		&	3 	&	5 	&	20 	&	-55 	&		&		&		&		\\							
Si-$\beta$	&	E1	&	37.12	(1)	&	37.123 	&	37.102 	&	37.01 	&		&		&		&		&		&		&	3 	&	-18 	&	-110 	&		&		&		&		&		\\							
Si-$\gamma$	&	E1	&	39.65	(1)	&	39.653 	&	39.642 	&	39.59 	&		&		&		&		&		&		&	3 	&	-8 	&	-60 	&		&		&		&		&		\\							
Si-$\delta$	&	E1	&	40.472	(4)	&	40.473 	&	40.460 	&	40.37 	&		&		&		&		&		&		&	1 	&	-12 	&	-102 	&		&		&		&		&		\\							
P-$\alpha$	&	M1	&	35.109	(2)	&	35.117 	&	35.115 	&	35.13 	&	35.032 	&		&		&		&		&		&	8 	&	6 	&	21 	&	-77 	&		&		&		&		\\							
P-$\beta$	&	M1	&	36.323	(2)	&	36.324 	&	36.328 	&	36.36 	&	36.313 	&		&		&		&		&		&	1 	&	5 	&	37 	&	-10 	&		&		&		&		\\							
P-$\gamma$	&	E1	&	38.268	(2)	&	38.280 	&	38.265 	&	38.19 	&	38.193 	&		&		&		&		&		&	12 	&	-3 	&	-78 	&	-75 	&		&		&		&		\\							
S-$\alpha$	&	E1	&	35.974	(2)	&	35.979 	&	35.963 	&	35.89 	&	35.947 	&		&		&		&		&		&	5 	&	-11 	&	-84 	&	-27 	&		&		&		&		\\							
S-$\beta$	&	E1	&	36.881	(3)	&	36.886 	&	36.871 	&	36.81 	&	36.784 	&		&		&		&		&		&	5 	&	-10 	&	-71 	&	-97 	&		&		&		&		\\							
S-$\gamma$	&	E1	&	38.072	(2)	&	38.074 	&	38.063 	&	38.03 	&	37.877 	&		&		&		&		&		&	2 	&	-9 	&	-42 	&	-195 	&		&		&		&		\\							
Cl-$\alpha$	&	E1	&	34.634	(2)	&	34.637 	&	34.628 	&	34.50 	&	34.188 	&		&		&		&	34.4 	&		&	3 	&	-6 	&	-134 	&	-446 	&		&		&		&	-234 	\\							
Cl-$\beta$	&	E1	&	35.043	(1)	&	35.047 	&	35.036 	&	34.91 	&	34.736 	&		&		&		&	34.8 	&		&	4 	&	-7 	&	-133 	&	-307 	&		&		&		&	-243 	\\							
Ar-$\alpha$	&	E1	&	33.340	(5)	&	33.350 	&	33.345 	&		&		&		&		&		&		&		&	10 	&	5 	&		&		&		&		&		&		\\							
Ar-$\beta$	&	E1	&	33.541	(1)	&	33.546 	&	33.544 	&		&		&		&		&		&		&		&	5 	&	3 	&		&		&		&		&		&		\\							
Ar-$\gamma$	&	E1	&	34.277	(5)	&	34.284 	&	34.280 	&		&		&		&		&		&		&		&	7 	&	3 	&		&		&		&		&		&		\\							
	&		&			&		&		&		&		&		&		&		&		&		&		&		&		&		&		&		&		&		\\							
Al-a	&	E1	&	63.18	(3)	&	63.251 	&	63.292 	&	63.89 	&	63.239 	&	63.23 	&	63.045 	&	63.27 	&		&		&	71 	&	112 	&	710 	&	59 	&	50 	&	-135 	&	90 	&		\\							
Al-b	&	E1	&	74.04	(2)	&	74.040 	&	74.015 	&	73.24 	&	74.177 	&	74.04 	&	74.076 	&	74.08 	&		&		&	0 	&	-25 	&	-800 	&	137 	&	0 	&	36 	&	40 	&		\\							
Si-a	&	E1	&	71.89	(3)	&	73.515 	&	73.528 	&	72.78 	&	72.847 	&		&		&		&		&		&	1625 	&	1638 	&	890 	&	957 	&		&		&		&		\\							
Si-b	&	E1	&	76.48	(3)	&	77.796 	&	77.838 	&	77.06 	&	77.344 	&		&		&		&		&		&	1316 	&	1358 	&	580 	&	864 	&		&		&		&		\\							
P-a	&	E1	&	53.96	(2)	&	54.008 	&	53.985 	&	53.81 	&	54.108 	&		&		&		&		&		&	48 	&	25 	&	-150 	&	148 	&		&		&		&		\\							
P-b	&	E1	&	76.07	(3)	&	76.267 	&	76.292 	&	75.49 	&	74.038 	&		&		&		&		&		&	197 	&	222 	&	-580 	&	-2032 	&		&		&		&		\\							
P-c	&	E1	&	100.42	(3)	&	100.39 	&	100.46 	&	101.2	&	99.362 	&		&		&		&		&		&	-26 	&	38 	&	780 	&	-1058 	&		&		&		&		\\							
S-a	&	E1	&	50.86	(2)	&	50.880 	&	50.857 	&	50.73 	&	51.182 	&		&		&		&		&		&	20 	&	-3 	&	-130 	&	322 	&		&		&		&		\\							
S-b	&	E1	&	52.80	(2)	&	52.828 	&	52.810 	&	52.70 	&	53.136 	&		&		&		&		&		&	28 	&	10 	&	-100 	&	336 	&		&		&		&		\\							
S-c	&	M1	&	164.44	(3)	&	164.50 	&	164.50 	&	165.3	&	165.19	&		&		&		&		&		&	63 	&	63 	&	860 	&	750 	&		&		&		&		\\							
Cl-a	&	E1	&	51.21	(2)	&	51.239 	&	51.235 	&	51.03 	&	51.262 	&		&		&		&	50.9 	&		&	29 	&	25 	&	-180 	&	52 	&		&		&		&	-310 	\\							
Cl-b	&	E1	&	52.54	(2)	&	52.570 	&	52.569 	&	52.37 	&	52.742 	&		&		&		&	52.2 	&		&	30 	&	29 	&	-170 	&	202 	&		&		&		&	-340 	\\							
Cl-c	&	M1	&	166.13	(3)	&	166.25 	&	166.22 	&		&		&		&		&		&		&		&	117 	&	90 	&		&		&		&		&		&		\\							
Cl-d	&	M1	&	169.11	(3)	&	169.22 	&	169.24 	&		&		&		&		&		&		&		&	113 	&	126 	&		&		&		&		&		&		\\							
Ar-a	&	E1	&	49.31	(2)	&	49.355 	&	49.366 	&		&		&		&		&		&		&		&	45 	&	56 	&		&		&		&		&		&		\\							
Ar-b	&	M1	&	140.98	(3)	&	141.02 	&	140.96 	&		&		&		&		&		&		&		&	43 	&	-23 	&		&		&		&		&		&		\\							
Ar-c	&	M1	&	147.85	(3)	&	147.86 	&	147.77 	&		&		&		&		&		&		&		&	7 	&	-76 	&		&		&		&		&		&		\\							
Ar-d	&	M1	&	153.59	(3)	&	153.68 	&	153.59 	&		&		&		&		&		&		&		&	86 	&	2 	&		&		&		&		&		&		\\							
Ar-e	&	M1	&	171.37	(3)	&	171.40 	&	171.33 	&		&		&		&		&		&		&		&	28 	&	-36 	&		&		&		&		&		&		\\							
Ar-f	&	M1	&	173.72	(3)	&	173.84 	&	173.82 	&		&		&		&		&		&		&		&	124 	&	98 	&		&		&		&		&		&		\\							

	\end{longtable}

\end{document}